\documentclass[floatfix]{elsart}

\usepackage{epsfig}
\usepackage{dcolumn}
\usepackage{amssymb}

\newcommand{\Etr} {$E^{12}_{\perp}$~}
\newcommand{\Mc} {$M_{\rm c}$~}    
\newcommand{\bbmax} {$b/b_{\rm max}$~}
\newcommand{\Zimf} {$Z>2$~}
\newcommand{\Zlcp} {$Z\leq2$~}
\newcommand{\XeSn} {\nuc{129}{Xe}+\nuc{\rm nat}{Sn} }
\newcommand{\AuAu} {\nuc{197}{Au}+\nuc{197}{Au} }
\newcommand{\XeSna} {\nuc{129}{Xe}+\nuc{\rm nat}{Sn}}
\newcommand{\AuAua} {\nuc{197}{Au}+\nuc{197}{Au}}

\newcommand{\vc} {$V_{\rm coll}$~}
\newcommand{\ac} {$\alpha_{\rm coll}$~}

\newcommand{\vol}[3]{\textbf{#1}\textrm{ (#3) #2}}
\newcommand{\NIM} {Nucl. Inst. and Meth. in Phys. Res.}
\newcommand{\NP} {Nucl. Phys.}
\newcommand{\PL} {Phys. Lett.}
\newcommand{\PR} {Phys. Rev.}
\newcommand{\PRL} {Phys. Rev. Lett.}
\newcommand{\RPP} {Rep. Prog. Phys.}
\newcommand{\ZP} {Z. Phys.}

\def\lsim{\mathrel{\rlap{
\lower4pt\hbox{\hskip-3pt$\sim$}}
    \raise1pt\hbox{$<$}}}     
\def\gsim{\mathrel{\rlap{
\lower4pt\hbox{\hskip-3pt$\sim$}}
    \raise1pt\hbox{$>$}}}     

\begin{document}

\begin{frontmatter}



\title{Statistical Multifragmentation of \\  Non-Spherical Expanding 
Sources in Central Heavy-Ion Collisions}
\author[GSI]{A. Le F{\`e}vre}, 
\author[GANIL]{M. P{\l}{}oszajczak}, 
\author[DUBNA]{V.D. Toneev},  
\author[GANIL]{G.~Auger},
\author[GSI]{M.L.~Begemann-Blaich},
\author[LPCC]{N.~Bellaize},
\author[GSI]{R.~Bittiger},
\author[LPCC]{F.~Bocage},
\author[IPNO]{B.~Borderie},
\author[LPCC]{R.~Bougault},
\author[GANIL]{B.~Bouriquet},
\author[SACLAY]{J.L.~Charvet},
\author[GANIL]{A.~Chbihi},
\author[SACLAY]{R.~Dayras},
\author[LPCC]{D.~Durand},
\author[GANIL]{J.D.~Frankland},
\author[IPNO,CNAM]{E.~Galichet},
\author[GSI]{D.~Gourio},
\author[IPNL]{D.~Guinet},
\author[GANIL]{S.~Hudan},
\author[LPCC]{B.~Hurst},
\author[IPNL]{P.~Lautesse},
\author[IPNO]{F.~Lavaud},
\author[SACLAY]{R.~Legrain \thanksref{deceased}},
\author[LPCC]{O.~Lopez},
\author[GSI,KRAKOW]{J. {\L}ukasik},
\author[GSI]{U.~Lynen},
\author[GSI]{W.F.J.~M{\"u}ller},
\author[SACLAY]{L.~Nalpas},
\author[GSI]{H.~Orth},
\author[IPNO]{E.~Plagnol},
\author[NAPOLI]{E.~Rosato},
\author[CATANIA]{A.~Saija},
\author[GSI]{C.~Schwarz},
\author[GSI]{C.~Sfienti},
\author[LPCC]{B.~Tamain},
\author[GSI]{W.~Trautmann},
\author[WARSAW]{A.~Trzci\'{n}ski},
\author[GSI]{K.~Turz{\'o}},
\author[LPCC]{E.~Vient},
\author[NAPOLI]{M.~Vigilante},
\author[SACLAY]{C.~Volant},
\author[WARSAW]{B.~Zwiegli\'{n}ski} and
\author[GSI,MOSCOW]{A.S.~Botvina} \\
(The INDRA and ALADIN Collaborations)

\address[GSI]{Gesellschaft f{\"u}r Schwerionenforschung mbH,
  D-64291 Darmstadt, Germany}
\address[GANIL]{GANIL, CEA and IN2P3-CNRS, F-14076 Caen, France}
\address[IPNO]{Institut de Physique Nucl{\'e}aire, IN2P3-CNRS and University,
  F-91406 Orsay, France}
\address[LPCC]{LPC Caen, IN2P3-CNRS, ENSICAEN and University, F-14050 Caen, France}
\address[SACLAY]{DAPNIA/SPhN, CEA/Saclay, F-91191 Gif-sur-Yvette, France}
\address[IPNL]{Institut de Physique Nucl{\'e}aire, IN2P3-CNRS and University,
  F-69622 Villeurbanne, France}
\address[NAPOLI]{Dipartimento di Scienze Fisiche e Sezione INFN,
  Univ. Federico II, I-80126 Napoli, Italy}
\address[CATANIA]{Dipartimento di Fisica dell' Universit\`{a} and INFN,
  I-95129 Catania, Italy}
\address[WARSAW]{A. So{\l}{}tan Institute for Nuclear Studies, Pl-00681 Warsaw,
  Poland}
\address[DUBNA]{Bogoliubov Laboratory of Theoretical Physics, 
  JINR, Dubna, 141980 Moscow Region, Russia}
\address[KRAKOW]{H. Niewodnicza\'{n}ski Institute of Nuclear Physics,
  Pl-31342 Krak\'{o}w, Poland}
\address[CNAM]{Conservatoire National des Arts et M{\'e}tiers, F-75141 Paris Cedex 03, France}
\address[MOSCOW]{Institute for Nuclear Research, 
  117312 Moscow, Russia}
\thanks[deceased]{Deceased.}

\begin{abstract}

  We study the anisotropy effects measured with INDRA at GSI in central 
  collisions of \XeSn at 50~A\,MeV and \AuAu at  
  60, 80, 100~A\,MeV incident energy.
  The microcanonical multifragmentation model 
  with non-spherical sources is used to simulate an incomplete shape 
  relaxation  of the multifragmenting system. This model is employed to
  interpret observed  anisotropic distributions in the fragment size and  
  mean kinetic energy. 
  The data can be well reproduced if an expanding prolate source
  aligned along the beam direction is assumed.
  An either non-Hubblean or non-isotropic radial expansion is 
  required to describe the fragment kinetic energies and their anisotropy.
  The qualitative similarity of the results for the studied reactions suggests 
  that the concept of a longitudinally elongated freeze-out configuration is 
  generally applicable for central collisions of heavy systems. The 
  deformation decreases slightly with increasing beam energy.
 
\end{abstract}

\end{frontmatter}

\section{Introduction}
\label{sec:intro}

Numerous transport calculations indicate that in central  
heavy-ion collisions at intermediate energies significant
compression and heating of nuclear matter  occur  in the initial 
stage of the reaction (see, e.g., \cite{suraud,aichelin91,peilert94,dani95}). 
The expansion of initially hot and compressed matter is thought to result 
in the development of dynamical instabilities leading to the 
multifragmentation, i.e. to the
formation of a large number of intermediate-mass fragments as it is 
observed experimentally \cite{bowman92,review}.
Traces of the pre-equilibrium processes in the disassembling source
and the flow energy initially stored in the compressed mode are expected to appear in the fragmentation patterns.
The formed composite system may, in particular, 
have a non-spherical or even a non-compact configuration which 
will affect the later stages of the decay process. 
Such configurations can be the result of a transparency, 
i.e. of an incomplete stopping of the colliding system, 
as reported for heavy-ion
 reactions at intermediate \cite{bao95,johnston,nebauer} 
and high \cite{rami} energies. The composite system disintegrates into
fragments before complete equilibrium has been attained.

Although the collision dynamics plays an essential role,
general features of the multifragmentation process, 
especially the fragment partitions, 
have been successfully described using the equilibrium 
statistical models \cite{gross,bondorf}.
In these models, the system is allowed to explore the phase space of possible partitions into light particles and fragments within an expanded volume. In their standard versions, a spherical breakup volume is used which causes the fragment distributions to be isotropic. 
While isotropy has been reported for fragmentations initiated with relativistic projectiles (see, e.g., \cite{schuettauf,beaulieu01,avdeyev02}) it is not necessarily a generic property of multifragmentation. In the heavy symmetric collision systems at intermediate energies studied in the present work, non-isotropic distributions are found to prevail in central collisions if standard methods are used for their identification \cite{nebauer,le_fevre,bouriquet,lavaud_berkeley}. 
A selection of subclasses of central events that fulfill the demand of isotropic emission patterns requires additional conditions for global event properties or for specific observables 
\cite{dagostino1,marie,bougault,marie98,frankland_GdU}.
The resulting cross sections are rather small.

Angular anisotropies in fragment distributions and kinetic energies
are usually considered as an indication for genuine non-equilibrium 
effects. Indeed, 
dynamical models are able to account for these anisotropies \cite{nebauer} but 
it is difficult with these models to assess the degree of equilibration that 
is otherwise reached at the fragmentation stage of the reaction 
\cite{nebauer99}. 
In a general statistical description, on the other hand, the 
shape degrees of freedom of the emitting sources will have to be 
included. 
Fluctuations around an average shape and isotropic orientations are expected in the unconstrained case while a preference for non-spherical shapes and non-isotropic orientations may result from dynamical effects. Spherical or near-spherical shapes can still be part of such dynamically constrained ensembles.

In the present paper, starting from the observed characteristics of central collisions, an effective deformation of the expanding fragmenting source is considered as a result of an incomplete shape relaxation. 
Apart from this dynamical imprint, the system is assumed to have attained equilibrium in all other degrees of freedom. The possibility of collective flow, decoupled from the statistical degrees of freedom, is furthermore included. 
In statistical analyses for the present and similar systems, an 
additional collective flow component has been shown to be essential for reproducing the fragment kinetic energies  
\cite{le_fevre,lavaud_berkeley,marie,bougault,jeong,steck96,raduta}. 
Customarily, it is 
superimposed without considering its potential correlation with the 
partition degree of freedom, an
assumption that finds its justification in 
the successful description of the experimental data that it permits 
\cite{bellaize}. 

In this framework, we shall analyze various
manifestations of anisotropy in fragment observables 
from central collisions of \XeSn at 50~A\,MeV and \AuAu at 60, 80 and
100~A\,MeV, measured at GSI with the INDRA multidetector.
The analysis is performed with the Metropolis Multifragmentation Monte-Carlo  
(MMMC) model \cite{gross} which has been extended to non-spherical (NS) 
sources~\cite{le_fevre}, a version referred to in the following as MMMC-NS model. The MMMC Statistical Model is based on the microcanonical ensemble  
and has found wide applications in the multifragmentation 
regime (see, e.g., \cite{gross,sneppen,marieluise}).

The \XeSn reaction has been previously studied at GANIL as part of the first 
INDRA campaign \cite{nebauer,marie,marie98,lukasik,gourio,plagnol,hudan}. 
Reactions of \AuAu have been studied at lower 
\cite{dagostino1,dagostino,huang97} as well as at higher ($\geq$ 100~A\,MeV)
incident energies \cite{tsang93,hsi94,reisdorf}. 
The present experiment covers the range 40 to 150~A\,MeV for \AuAua, 
and the new results obtained for peripheral collisions have been 
reported \cite{lukasik02,lukasik03}.
A study of central collisions and their analysis within the statistical multifragmentation model (SMM, Ref.~\cite{bondorf}) 
has been performed by Lavaud \textit{et al.}
for the range 40 to 100~A\,MeV~\cite{lavaud_berkeley,lavaud}.

The paper is organized as follows: experimental details and the criteria 
used for selecting central collisions are presented in 
Sect.~\ref{sec:data_sel}.
Central collisions of \XeSn at 50~A\,MeV incident energy are chosen to 
illustrate the analysis procedure (Sect.~\ref{sec:XeSn}), and their 
general characteristics are given in Sect~\ref{sec:s_size_eexc}. 
Using the MMMC-NS model, we first determine the general properties 
of the source, its mass and excitation energy (Sect.~\ref{sec:s_source}), 
and the strength of the collective radial flow (Sect.~\ref{sec:s_flow}). 
We then detail the protocol developed to investigate independently 
the geometrical properties of the source (Sect.~\ref{sec:s_elong}) and the
velocity profile of the collective radial flow  
(Sect.~\ref{sec:f_dim}). Standard event-shape variables are finally used to
confirm the validity of the derived conclusions (Sect.~\ref{sec:e_shape_var}).
To investigate the dependence of the anisotropy effects on the mass of the system and on the bombarding energy, we apply the same analysis to the heavier
system \AuAu at the similar beam energy 60~A\,MeV (Sect. \ref{sec:AuAu60})
and at the higher energies 80 and 100~A\,MeV (Sect.~\ref{sec:AuAu_rest}). 
Finally, in Sect. \ref{sec:conclusions} we summarize and discuss the main 
results of the paper.

\section{Experimental details and data selection}
\label{sec:data_sel}

The experimental data have been obtained during the INDRA campaign 
performed at the GSI laboratory in 1998 and 1999. 
Beams of \nuc{129}{Xe} and \nuc{197}{Au}, delivered by the heavy ion 
synchrotron SIS were directed onto $^{\rm nat}$Sn and Au targets with 
areal densities of 1.05~mg/cm$^{2}$ and 2.0~mg/cm$^{2}$, respectively.
The reaction products were detected and identified with the $4\pi$ multidetector
INDRA  described in detail in
\cite{pouthas,pouthas_electron}. For the present campaign, the 
first ring of the multidetector, covering the polar angles between 
2$^{\circ}$ and 3$^{\circ}$, has been modified in order to improve the energy
and charge resolution and to permit the mass identification of light products 
in the range $Z$ = 1 to 4. Without changes of the ring geometry, the previously
used phoswich detectors have been replaced by 12 Si-CsI(Tl) telescopes, each 
consisting of a 300-$\mu$m Si detector and a 14-cm-long CsI(Tl) scintillator 
viewed by a photomultiplier. 
These new detectors have the same characteristics as 
those used in the neighbouring INDRA rings. 

In order to verify that 
the  SIS beams  are properly focused into the 
12-mm-diameter entrance hole of INDRA, annular plastic detectors 
were placed upstream of the INDRA detection system, and their signals recorded for vetoing halo events in the off-line analysis. Measurements with empty 
target frames have also been performed. The event triggering condition was set to a hit multiplicity greater than 2 over all CsI detectors. This was sufficient to exclude elastic scattering and to suppress other background generated by cosmic rays and natural or induced radioactivity near the detectors. 

The energy calibration of \Zlcp particles for the 192 CsI(Tl) detectors of
the 9 forward-most rings of INDRA has primarily been derived from a
detector-by-detector comparison of spectra obtained in the reaction
\XeSn at 50~A\,MeV with those measured for the same reaction at GANIL.
There, calibration data had been
obtained by scattering a variety of primary and  secondary beams from thick
targets \cite{marie95}. 
The remaining 144 CsI(Tl) detectors of rings 10-17 have been calibrated using
the standard method \cite{pouthas} based on the silicon calibration
telescopes (one per ring): the spectra of \Zlcp particles
obtained for each scintillator have been adjusted to those measured
simultaneously by the telescopes during an experiment.
The validity of these calibration procedures has been verified for part of the 
detector by analyzing the spectra of recoil protons from $^{1}$H + $^{12}$C 
scattering, obtained with $^{12}$C beams of 30~A\,MeV and a polyethylene target~\cite{trzcinski}.

In order to calibrate the energies of heavier fragments (from Li to Au) with
the use of the calibration parameters obtained for $\alpha$-particles for all
INDRA modules (rings 1-17), the light emission from the CsI(Tl) scintillators
has been parameterized according to Ref. \cite{parlog}, and the additional
$Z$ dependence of the parameters was obtained by comparing the experimental
$\Delta$E-E maps to predictions of the range-energy tables \cite{hubert}.

\begin{figure}
\begin{center}
  \epsfysize=8.0 cm
  \epsffile{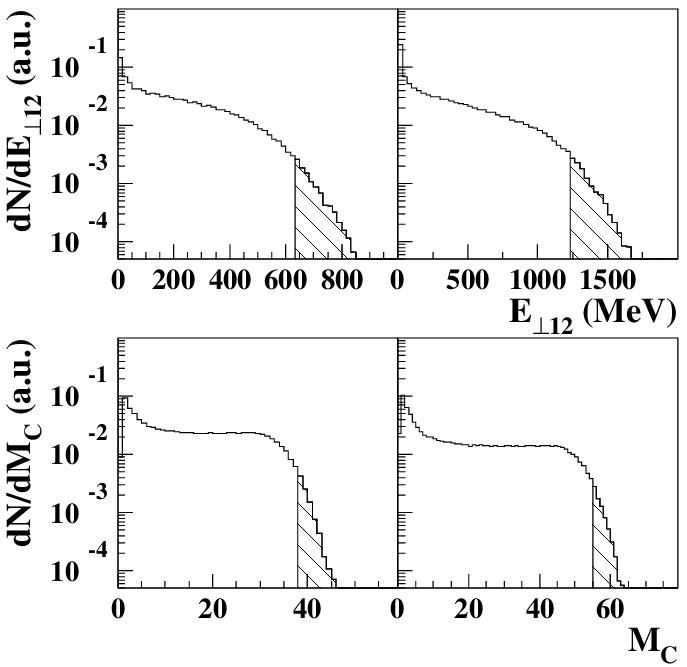}

  \caption{Top panels: Distributions of the total transverse energy of 
    light charged particles
    (\Zlcp) for the reactions \XeSn at 50~A\,MeV (left)
    and \AuAu at 60~A\,MeV (right). The hatched areas indicate 
    central events with \bbmax $\leq0.1$ selected according to  this 
    observable (E$^{12}_{\perp} \geq 637$~MeV and E$^{12}_{\perp} \geq 1256$~MeV, 
    respectively). 
    Bottom panels: Corresponding distributions of charged particle 
    multiplicity \Mc for the same two reactions. 
    The reduced impact parameter \bbmax $\leq0.1$
    corresponds to \Mc $\geq38$ and \Mc $\geq55$, 
    respectively (hatched areas).}
    \label{fig:mc_etrans12_2}
\end{center}
\end{figure}

The total transverse energy \Etr of light charged particles (\Zlcp) which has 
been used in previous studies \cite{nebauer,lukasik,plagnol,lukasik02} 
but also the multiplicity \Mc of charged particles were chosen to select
central collisions. 
Inclusive spectra of these quantities for the reactions \XeSn at 50~A\,MeV 
and \AuAu at 60~A\,MeV are shown in Fig.~\ref{fig:mc_etrans12_2}.
The hatched areas indicate the groups of central collisions accepted in 
the present analysis. They correspond to 1\% of the recorded reaction cross 
section or to reduced impact parameters \bbmax $\leq0.1$ if a monotonic 
relation with the impact parameter is assumed. 
It is obvious, in particular from the \Mc spectra, that these high thresholds 
favour the tails of the distributions which may be partly caused by  
the fluctuation widths of \Etr and \Mc in central collisions rather than
by a significant variation of the impact parameter up to the highest values 
of these variables.
 
\begin{figure}
\begin{center}
  \epsfysize=5.0 cm
  \epsffile{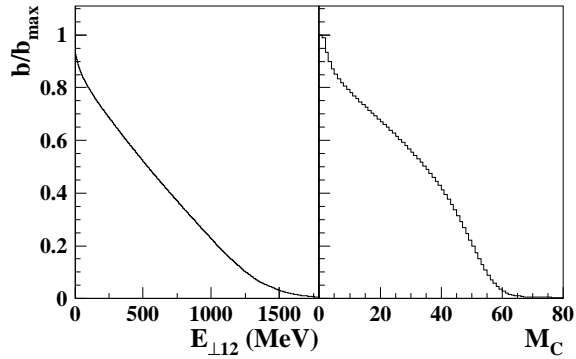}
  \caption{
    Reduced impact parameter \bbmax as a function of the total 
    transverse energy of light charged particles with \Zlcp (left) 
    and of the multiplicity of charged particles (right) for
    the reaction \AuAu at 60~A\,MeV. The trigger condition of at least
    three hits is 
    only partially reflected in \Mc because hits may be generated also
    by gamma rays and neutrons.
  }
  \label{fig:bbmax_r4227}
\end{center}
\end{figure}

The monotonic relation between \Etr and 
the reduced impact parameter \bbmax  resulting from
the geometrical prescription \cite{cavata} is linear 
to a good approximation, with \bbmax decreasing with increasing $E^{12}_{\perp}$.
This is illustrated in Fig.~\ref{fig:bbmax_r4227} (left panel)
for the system \AuAu at 60~A\,MeV.
The relation between \Mc and \bbmax for this reaction 
(Fig.~\ref{fig:bbmax_r4227}, right panel)
exhibits a change of slope near \Mc = 40 beyond which \bbmax varies much more
rapidly with $M_{\rm c}$. A similar behavior is not apparent in the 
correlation of \Etr with $b/b_{\rm max}$.

\begin{figure}
\begin{center}
  \epsfysize=6.5 cm
  \epsffile{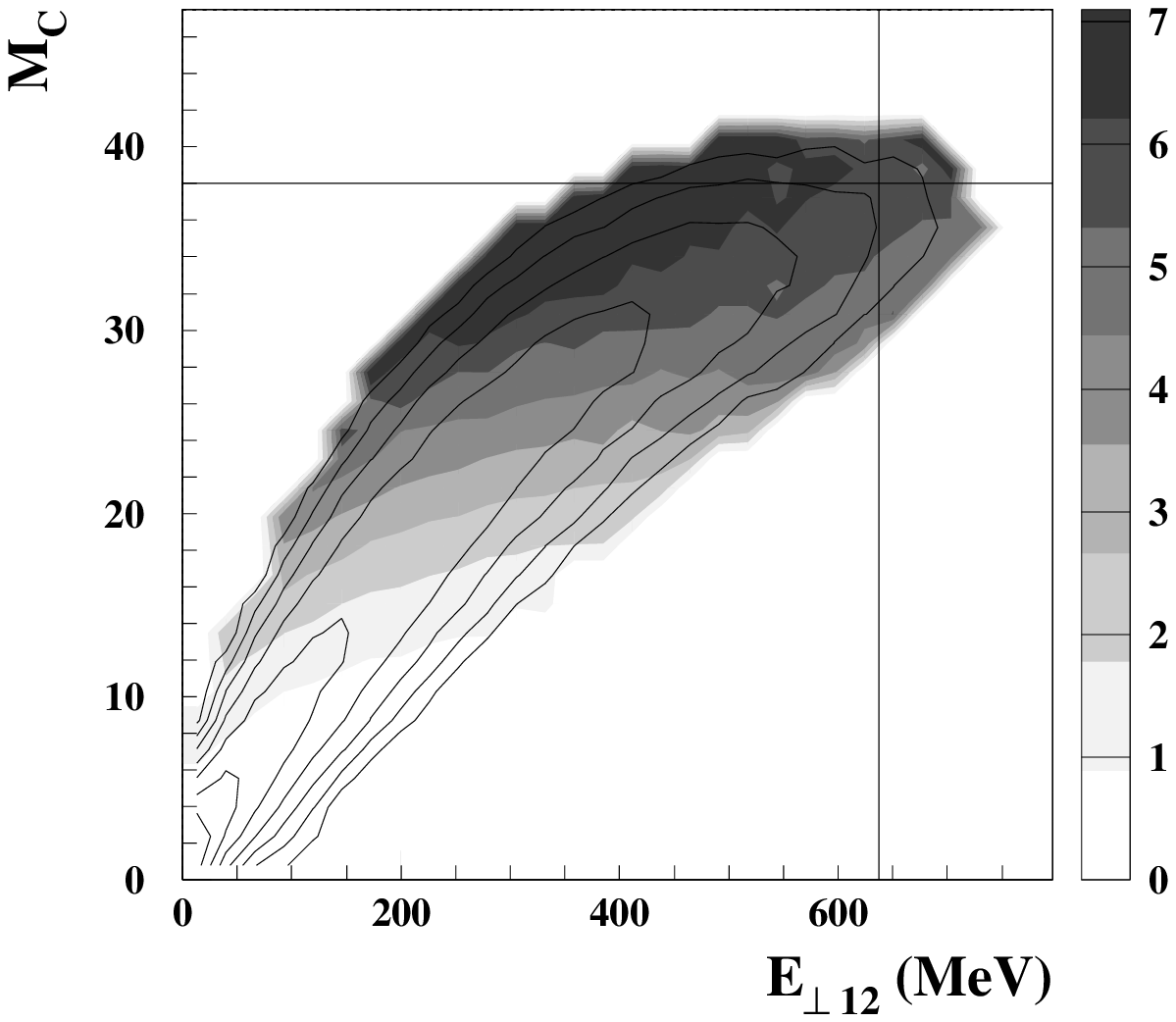}
  \epsfysize=6.5 cm
  \epsffile{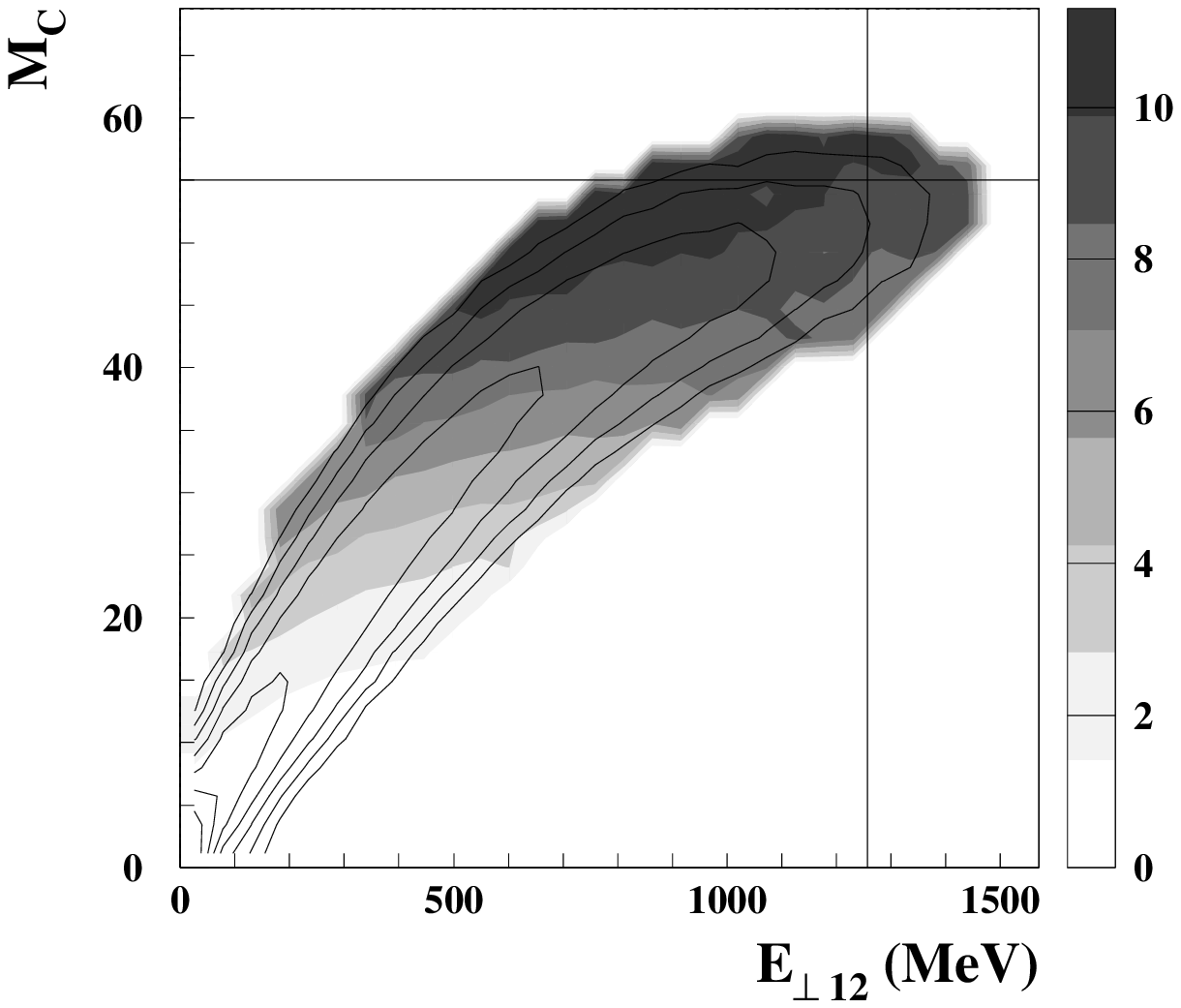}
  \caption{
    Double-differential reaction cross section (logarithmic contour 
    lines) and mean 
    multiplicity of fragments with \Zimf (shadings) as functions 
    of the multiplicity of charged particles \Mc and the total 
    transverse energy
    \Etr of light charged particles (\Zlcp) for \XeSn collisions at
    50~A\,MeV (left) and \AuAu at 60~A\,MeV (right). 
    The horizontal and vertical lines indicate the minimum values 
    of \Mc and \Etr that have been used to select central events in 
    these reactions.
    }
    \label{fig:mc_etrans12} 
\end{center}
\end{figure}

The correlation between the two variables \Etr and \Mc is shown in 
Fig.~\ref{fig:mc_etrans12} for the same two systems \XeSn at 50 A\,MeV and 
\AuAu at 60 A\,MeV. It is linear at small values of these observables, 
i.e. for peripheral collisions. For the most violent collisions we 
observe a tendency of \Mc to saturate.
The multiplicity of intermediate-mass fragments ($Z>2$), represented by the 
shadings in the figure, is more strongly correlated with \Mc than with 
$E^{12}_{\perp}$. For a given total multiplicity \Mc or multiplicity of fragments, the 
fluctuations of \Etr are rather large.
There is even the indication of a weak anti-correlation  
between the multiplicity of fragments and $E^{12}_{\perp}$; the largest number
of fragments is observed for the region of largest \Mc but, at a given bin of
$M_{\rm c}$, at the smallest $E^{12}_{\perp}$. This observation may reflect the fact 
that the production of fragments is governed by the amount of energy that is 
left available in the system and, apparently, subject to a competition 
with the energy used for light-particle emission. The figure, furthermore, 
shows that the central event classes selected with the two criteria are to 
a large extent mutually exclusive. 

\section{Central collisions of \XeSn at 50~A\,MeV} \label{sec:XeSn}
\subsection{General characteristics} \label{sec:s_size_eexc}

The global characteristics of central collisions of 
\XeSn at 50~A\,MeV, identified on the basis of either \Etr or $M_{\rm c}$, 
are shown in Table~\ref{tab:table1}. 
They are largely independent of the chosen selection criterion, except for 
the charged-particle and fragment multiplicities which are higher for the \Mc selected event class. 
The difference of about 20\% reflects the autocorrelation introduced by 
gating on the multiplicity itself (Fig.~\ref{fig:mc_etrans12}).
In order to illustrate the effect of the incomplete acceptance of the 
detector, the same observables are also evaluated for the subclass of 
events for which at least 80\% of the system charge has been recovered 
in the form of detected charged particles and fragments 
(values given in parentheses).
This additional condition has a slightly larger effect on the \Etr selected 
than on the \Mc selected event class.
A high multiplicity of charged particles, $M_{\rm c} > 30$, and a 
fraction of about 50\% of the system charge bound in fragments characterize 
this multifragmentation mode. Correspondingly, even the biggest fragments 
of the partition ($Z_{\rm max1}\approx12$ to 15) are relatively small, 
with the next largest fragments rapidly decreasing in charge. 
The sum of the transverse kinetic energies of charged particles reaches up to 
1.0~GeV, of which about two thirds are carried away by light charged particles with \Zlcp (Fig.~\ref{fig:mc_etrans12_2}). This dissipation amounts to about 
one third of a total of 3.1~GeV incident energy in the center of mass and, 
even though neutrons and the separation energies are not included, may be 
taken as a first hint that the projectile is not completely stopped in 
these most central collisions.

\begin{table}
  \caption{
    Characteristics of central collisions of \XeSn 
    at 50~A\,MeV, selected by the  \Mc or \Etr cuts,
    and MMMC-NS predictions with the prolate source adjusted 
    to the \Etr selected experimental events (for the source parameters
    see Table~\protect\ref{tab:table2} ). 
    Mean values are given for the observables (from top to bottom) 
    multiplicity \Mc of charged particles, multiplicity $M_{\rm frag}$ of
    fragments with \Zimf, the atomic numbers $Z_{\rm max}$ of the  
    largest ($Z_{\rm max1}$) and of the second and third largest fragments
    ($Z_{\rm max2}, Z_{\rm max3}$), sphericity $s$
    and coplanarity $c$ (Sect.~\protect\ref{sec:e_shape_var}), 
    sum $\Sigma Z$ of the fragment charges, and their kinetic energies $\Sigma E_{\rm kin}$
    in the center of mass (in units of MeV), summed over \Zimf and $Z>5$.
    The values in parentheses are obtained 
    for the event class with 80\% completeness in charge detection. 
    Using the fact that the system is nearly symmetric, the multiplicities and     the sums over charge and kinetic energies are obtained by doubling 
    the mean values obtained for the forward hemisphere 
    ($\theta_{\rm cm}\leq90^{\circ}$) where
    the acceptance for fragments is higher. 
    }
    \label{tab:table1}
\begin{center}
  \begin{tabular} {l c c c}
    \hline
    Observables & \multicolumn{2}{c}{Experiment} & MMMC-NS\\  
    \cline{2-3}
    & \Etr selection & \Mc selection & \\
    \hline
    \Mc & 33 (34) & 39 (39) & 20 (21) \\
    $M_{\rm frag}(Z>2)$ & 5.3 (6.3) & 6.5 (6.9) & 5.9 (6.6) \\
    $Z_{\rm max1}$ & 13.3 (15.6) & 12.3 (13.1) & 13.7 (15.0) \\
    $Z_{\rm max2}$ & 7.4 (9.9) & 7.8 (8.5) & 8.0 (8.7) \\
    $Z_{\rm max3}$ & 4.3 (6.6) & 5.4 (6.2) & 6.1 (6.6) \\
    $s$ & 0.49 (0.53) & 0.55 (0.56) & 0.56 (0.57) \\
    $c$ & 0.15 (0.14) & 0.14 (0.14) & 0.14 (0.13) \\
    $\Sigma Z (Z>2)$ & 39 (50) & 42 (46) & 44 (50) \\
    $\Sigma E_{\rm kin} (Z>2)$  & 400 (491) & 418 (456) & 407 (468) \\
    $\Sigma E_{\rm kin} (Z>5)$  & 252 (336) & 256 (292) & 256 (305) \\
    \hline
  \end{tabular}
\end{center}
\end{table} 

\subsection{Global parameters of the statistical source} \label{sec:s_source}

It is well established for this symmetric reaction that
a heavy, hot and expanding composite system is formed in central collisions
\cite{nebauer,marie,bougault,marie98,bellaize,hudan}.
Its properties have been previously determined by requiring 
the Statistical Multifragmentation Model (SMM, Ref.~\cite{bondorf}) 
to reproduce the partitions and the overall mean kinetic energies of 
fragments~\cite{bougault}. The obtained source characteristics
at freeze-out (at a density of 1/3 of the normal density $\rho_0$) were 
a total charge $Z_{\rm S}=78$, an excitation 
energy of 7~A\,MeV, and a collective radial flow of 2.2~A\,MeV in addition to the thermal and Coulomb components of the kinetic energies.  According to calculations with the BNV
 transport model~\cite{bougault}, these parameters correspond
to those of an expanding source in the spinodal region.

\begin{figure}
\begin{center}
  \epsfysize=6.0 cm
  \epsffile{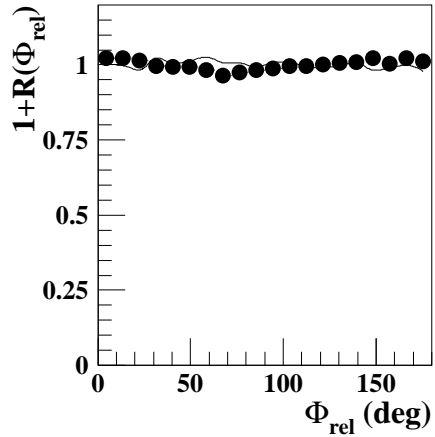}
  \caption{
    Correlation functions in relative azimuthal angle between 
    alpha particles.
    The symbols are the experimental data for central \XeSn collisions at
    50~A\,MeV and the line represents the MMMC-NS model
    prediction with $40~\hbar$ angular momentum. 
    }
    \label{fig:alpha_correl} 
\end{center}
\end{figure}

This analysis has been repeated with the present data and within the 
MMMC-NS statistical approach \cite{le_fevre}. For this purpose,
the code was used in its standard configuration 
with a spherical source and with the fixed standard value $R=2.2A^{1/3}$ fm
for the freeze-out radius, corresponding to 1/6 of the normal nuclear 
density~\cite{sneppen,schapiro,schapiro2}. 
Collective radial (see below) and angular motions were also included. 
The chosen angular momentum of $40~\hbar$, 
perpendicular to the beam axis, represents an upper limit for this 
reaction for the lowest range of impact parameters~\cite{steckmeyer}.
The correlation function in the relative azimuthal angle 
between alpha particles is flat (Fig.~\ref{fig:alpha_correl}), indicating 
that the transfer of angular momentum to the formed composite system is too small to permit a pronounced reaction plane to be discerned \cite{tsang}.
In the MMMC-NS calculations, the inclusion of $40~\hbar$ angular 
momentum does not significantly alter  
the fragment partitions, kinetic energies or angular distributions.
The calculated alpha-particle correlation function 
remains also flat in accordance with the measurement
(Fig.~\ref{fig:alpha_correl}).
 
In all comparisons with the experimental data,
the produced MMMC-NS events have been 
filtered using a software replica of the experimental apparatus.
It takes into account the major properties of the INDRA
multidetector, i.e. the geometrical acceptance and the energy thresholds and 
high-energy cutoffs. 
In addition, the effects of the multihit treatment in the off-line analysis,
based on consistency tests of the recorded $\Delta E$ and $E$ signals
(so-called coherency), are also simulated. 
Applying this filtering procedure to the MMMC-NS generated events, 
 acceptances of $\approx80\%$ and $\approx70\%$ for  fragments
 ($Z>2$) and  light charged particles ($Z=1,2$) were obtained, respectively. 
They are caused by the geometrical acceptance of INDRA ($\approx90\%$), 
by thresholds, and by
multihits that cannot be resolved with the coherency
 analysis, in the majority multihits of light charged particles in a single detector module. Accordingly, the fragment multiplicity ($Z>2$) for \Etr 
selected central 
events, corrected for instrumental effects, is 6.6 ($>6.9$ for 80\% 
completeness, cf. Table~\ref{tab:table1}).

\begin{figure}
\begin{center}
  \epsfysize=6.0 cm
  \epsffile{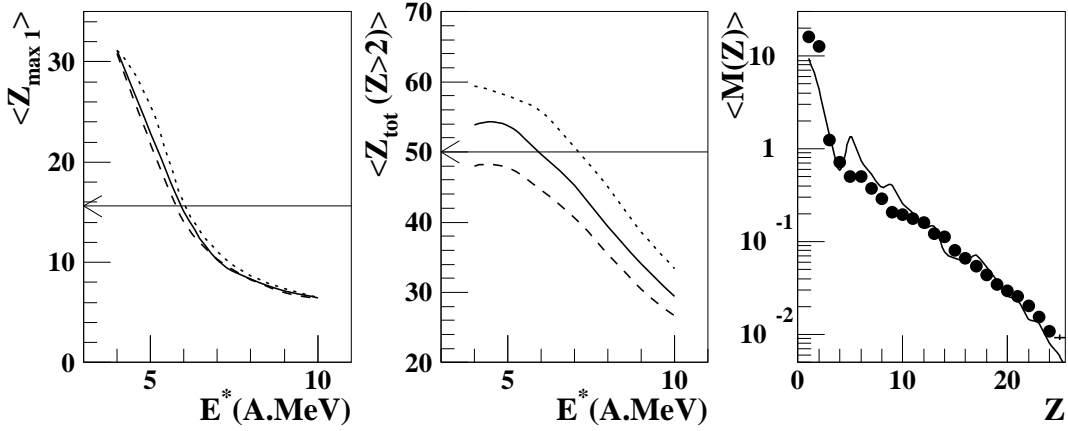}
  \caption{
    Left: Mean value of the largest fragment charge within an event 
    ($Z_{\rm max1}$) as a function of the excitation energy for MMMC-NS
    predictions with different source size: $Z_{\rm S}=70$ (dashed line),
    $Z_{\rm S}=79$ (full line) and $Z_{\rm S}=90$ (dotted line). 
    The arrow points to the experimental value for 
    central \XeSn at 50~A\,MeV selected using E$^{12}_{\perp}$.
    Middle: Mean total charge of fragments (\Zimf) for the same sources
    as in the left panel and the experimental value for the same 
    reaction. 
    Right: Mean measured multiplicity of fragments with charge $Z$ 
    (circles). 
    The line is the MMMC-NS prediction with the source size
    $Z_{\rm S}=79$ and  excitation energy $E^{*}=6$~A\,MeV. In all cases,
    only events having a detected  total charge $Z_{\rm tot}\geq80\%$ of
    the initial system charge have been included.  
    }
    \label{fig:mmmc_systematics} 
\end{center}
\end{figure}

The source parameters were determined by adjusting the calculations in a step-wise procedure to the measured mean characteristics of the fragment partitions.
The obtained results were found to be independent of whether the additional 
condition of $80\%$ charge detection, in both data and filtered calculations,
 had been requested. 
In a first step, the excitation energy of the source, $E^{*}=6.0$~A\,MeV, has 
been determined by requiring the mean value of the largest
fragment charge $Z_{\rm max1}$ of the event to be reproduced. 
This can be done independently of
the mass of the source since $E^{*}$ depends mainly  on $Z_{\rm max1}$ 
and is only weakly sensitive to the total charge of the source $Z_{\rm S}$
(Fig.~\ref{fig:mmmc_systematics}, left panel).
This excitation energy 
 corresponds to a mean thermodynamical temperature $T=5.6$~MeV. 
In a second step, with $E^{*}$ being fixed, the source charge $Z_{\rm S}$ is
determined by adjusting the total detected charge of fragments
$\Sigma Z (Z>2)$ to its experimental mean value. 
This observable depends strongly on both the source mass and the excitation energy (Fig.~\ref{fig:mmmc_systematics}, middle panel). 
For $E^{*}=6.0$~A\,MeV, a value $Z_{\rm S}=79$ is obtained for the charge of the equilibrium source of fragments. 
The assumption that the mean $N/Z$ ratio of the colliding nuclei 
is preserved yields $A_{\rm S}=188$ for the mass.  
These values correspond to $76\%$ of the total system. 
The simulations performed in Ref.~\cite{le_fevre} for an excited
gold nucleus with $Z_{\rm S}=79$ and $A_{\rm S}=197$ yield practically identical results for the observables studied here, in spite of the larger neutron number.
The resulting 
charge distribution of fragments ($Z>2$) reproduces well the 
experimental data (Fig.~\ref{fig:mmmc_systematics}, right panel).
The difference observed for light particles (\Zlcp) is presumably caused by
pre-equilibrium emissions during early phases of the reaction
(cf. Sec.~\ref{sec:s_elong}). 

While the system mass obtained from the MMMC analysis 
agrees with the SMM estimate~\cite{bougault}, 
the excitation energy is slightly lower. This can at least  
partly be traced back to the freeze-out volume which in the 
MMMC-NS model has twice the value used in the SMM. 
The Coulomb energy is correspondingly lower, by about 0.8 A\,MeV for the 
present system, so that the thermal components of the excitation energy 
which determine the fragment partitions are about equal. 

\subsection{Collective radial flow} \label{sec:s_flow}

A collective radial flow of the constituents of the source
was incorporated into the MMMC-NS code with the parameterization
\begin{equation} \label{eq:vcoll}
\overrightarrow{v}_{coll}(\vec r)=V_{coll} 
 \left(\frac{r}{R_0}\right)^{\displaystyle \alpha_{coll}} \ \left(\frac{\vec r}{r}\right)
\end{equation} 
where $\vec r$ is the position vector in coordinate 
space with respect to the center of mass, and 
$V_{\rm coll}$, $R_0$ and \ac denote the strength, the scaling radius and the 
exponent of the flow profile, respectively. 
$R_0$ is set to be equal to the radius of the system at normal density.
To obtain satisfactory descriptions of the observed anisotropies, it was found necessary to either permit \ac to differ from unity, i.e. to deviate from the
standard Hubblean self-similar flow profile \cite{le_fevre,raduta,morawetz},
or to consider non-isotropic strength parameters $V_{\rm coll} (\theta_{\rm cm})$ 
\cite{lavaud_berkeley,lavaud}.   
The collective energy is not included in the energy balance, and the flow 
is thus decoupled from the equilibrium.
This assumption is, e.g., justified in a quasi-static-expansion approximation in 
which the expansion is slow as compared to the equilibration time. 

The strength parameter \vc is determined from the requirement that 
the sum of the kinetic energies of fragments with $Z>5$, $\Sigma E_{\rm kin} (Z>5)$ 
as given in Table~\ref{tab:table1}, 
is reproduced. In the cases with isotropic $V_{\rm coll}$, a flow exponent \ac = 2
has been used since the observed anisotropies in momentum space are best 
accounted for with this value (Sect.~\ref{sec:f_dim}). This is not 
crucial for the spherical source for which the result for \vc is nearly 
independent of the choice made for $\alpha_{\rm coll}$, a consequence of the specific 
parameterization Eq.~\ref{eq:vcoll} and the chosen scaling radius $R_0$.
The obtained value \vc $=0.07 c$ corresponds to 
a mean collective energy $<E_{\rm coll}>$ = 2.4~A\,MeV. This result is close to the 
values 2.2~A\,MeV from the SMM analysis~\cite{bougault} and 2.0~A\,MeV 
obtained with the phenomenological model used in \cite{marie}.
The spherical-source characteristics are summarized in Table~\ref{tab:table2}.

From the sum of the collective energy and the thermal excitation 
energy, we deduce a total excitation energy of 8.4~A\,MeV at 
freeze-out.
This value is considerably smaller than the 12~A\,MeV that have 
been extracted experimentally for a small subset of central collisions of the same system with a calorimetric method \cite{marie}. A difference of several MeV per nucleon between the excitation energies needed to describe the fragmentation within statistical models and those deduced from summing up the breakup Q-values and kinetic energies has also been observed for other types of reactions and ascribed to preequilibrium emissions \cite{schuettauf}. Also in the present case the emission of light charged particles is underestimated by the MMMC 
model (Table \ref{tab:table1} and Fig. \ref{fig:mmmc_systematics}, right panel). Part of the difference may also be caused by the selection of a specific event sample with high sphericity that was chosen in \cite{marie}.

\begin{table}
  \caption{
    Source characteristics for \XeSn at 50~A\,MeV used in the MMMC-NS 
    calculations: charge $Z_{\rm S}$, mass $A_{\rm S}$ and excitation energy $E^*$ 
    of the source,  
    transversal-to-longitudinal elongation ${\mathcal R}$ in coordinate space,
    collective radial flow parameters \vc (strength) and \ac 
    (exponent), and resulting mean kinetic energy of the radial flow.
    The elongation of the ellipsoidal flow in velocity space is 0.30:1 and
    0.70:1 for the spherical (Sphere+e.f.) and prolate (Prolate+e.f.) sources,
    respectively.
    In all cases, the angular momentum perpendicular to the beam axis is  
    $40~\hbar$, and the reduced freeze-out radius is $r_{0}=2.2$~fm.
    }
    \label{tab:table2}
    \begin{center}
    \begin{tabular} {l c c c c c}
      \hline
      Source shape & Prolate & Sphere & Oblate & Sphere+e.f. & Prolate+e.f. \\
      \hline
      $Z_{\rm S}$ & 79 & 79 & 79 & 79 & 79 \\
      $A_{\rm S}$ & 188 & 188 & 188 & 188 & 188 \\
      $E^*$ (A\,MeV) & 6.0 & 6.0 & 6.0 & 6.0 & 6.0 \\
      ${\mathcal R}$ & 0.70:1 & 1:1 & $1.67:1$ & 1:1 & 0.70:1 \\
      \vc (c) & 0.06 & 0.07 & 0.05 & 0.07 & 0.06 \\
      \ac & 2 & 2 & 2 & 1 & 1 \\
      $<E_{\rm coll}>$ (A\,MeV) & 2.3 & 2.4 & 1.8 & 1.9 & 2.1 \\
      \hline 
    \end{tabular}
  \end{center}
\end{table}

The global event-shape observables sphericity and coplanarity, deduced from 
the three-dimensional tensor of fragment momenta \cite{cugnon} and discussed 
in detail in Sect. \ref{sec:e_shape_var}, are included in 
Table~\ref{tab:table1}. The experimental mean sphericity determined for 
the \Etr selected event class is $<s>$ = 0.49 and, thus, considerably 
smaller than the prediction $<s>$ = 0.71 obtained, after filtering, with the
MMMC-NS code in the spherical-source limit (note that the limiting values 
$<s>$ = 1.0 and $<c>$ = 0.0 cannot be reached with a finite number of 
fragments).
The coplanarity is small which, for non-spherical shapes, is expected for 
prolate deformations with approximately azimuthal symmetry.
The spherical-source model, while adequately accounting for the partition degrees of freedom, cannot reproduce the source shape in momentum space. 
As will be shown in the following, the indicated prolate source deformation 
is confirmed
by the results obtained for other observables which reflect the angular
dependence of the masses and the kinetic energies of the fragments. 
These anisotropies 
indicate that the system has not completely relaxed and that part
of its total energy is bound in other degrees of freedom  
and, hence, not available for the thermal equilibration.

\subsection{Anisotropy of heavy-fragment distribution 
and the spatial elongation of the source} 
\label{sec:s_elong}
 
In the MMMC-NS statistical approach \cite{le_fevre}, 
the fragment partitions are accomodated within a given source volume.
Besides the spherical shape, also prolate and oblate source geometries are 
included here, i.e. elongated and compressed
ellipsoids in coordinate space, aligned with respect to the beam axis.
These three kinds of sources are meant to represent
possible dynamical scenarios for the source formation prior to freeze-out. 
For the spherical and prolate sources, besides isotropic radial flow, also the
possibility of ellipsoidal radial flow is considered. 
It is generated by weighting the parameter \vc (Eq.~\ref{eq:vcoll}) 
with a geometrical function that describes an ellipsoid in momentum space
elongated along the beam axis.
The deformation is adjusted so as to 
reproduce the experimental angular dependence of the fragment
kinetic energies (see Sect.~\ref{sec:f_dim}). 
 
\begin{figure}
\begin{center}
  \epsfysize=6.0 cm      
  \epsffile{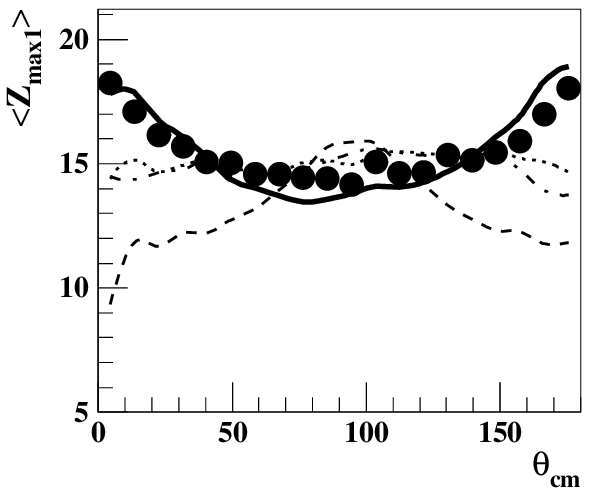}
  \caption{
    Mean measured value of the largest fragment charge within an event 
    ($Z_{\rm max1}$) as a function of the polar angle in the 
    center-of-mass, $\theta_{\rm cm}$, for central collisions of 
    \XeSn at 50~A\,MeV selected using \Etr (circles) with the 
    requirement $Z_{\rm tot}\geq80\%$. MMMC-NS model
    predictions are given for a prolate (full line), a spherical 
    (dotted) and an oblate (dashed) source, and for a spherical
    source with non-isotropic radial flow (dashed-dotted line). 
    }
    \label{fig:Zmax_vs_theta} 
\end{center}
\end{figure}

We shall now analyze various observables related to the charge distributions 
and anisotropies and test their sensitivity to  
the source elongation in coordinate space. 
The charge of the largest fragment of a partition has a
mean value $\langle Z_{\rm max1} \rangle = 15.6$ (Table \ref{tab:table1}) 
but is on average $\approx20\%$ smaller if 
the heaviest fragments are emitted sidewards ($\theta_{\rm cm}\approx90^{\circ}$) than when they are
emitted in forward or backward directions (Fig.~\ref{fig:Zmax_vs_theta}). 
Already this first experimental observation can only
be reproduced with the assumption of a prolate source shape
but not with any of the other alternatives.
The best fit to the data is obtained with a spatial elongation 
given by the transversal-to-longitudinal ratio ${\mathcal R}=0.70:1$, 
determined with a statistical accuracy of $\pm3\%$. 
This corresponds to a deformation
parameter~\cite{bohr_mottelson}
\begin{displaymath}
\delta = \frac{3}{2} \ \frac{1-\mathcal{R}^{2}}{1+2\mathcal{R}^{2}} = 0.386.
\end{displaymath}
In the following, the elongation of the prolate source will be kept fixed.
For the oblate source, further used for illustration purposes, 
an elongation 1.67:1 has been arbitrarily chosen. 
Moderate deformations of this magnitude do not significantly modify 
the partitions. Vice versa, the global source parameters mass, 
excitation energy and the strength of the collective flow, as determined
for the various types of sources, are approximately 
the same (Table~\ref{tab:table2}).

\begin{figure}
\begin{center}
  \epsfysize=5.0 cm
  \epsffile{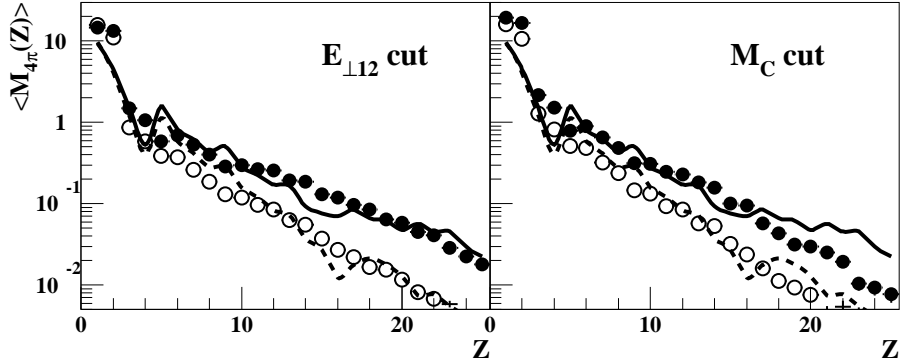}
  \caption{
    Mean multiplicity of fragment charges, normalized to a
    solid angle of $4\pi$, with selection of central events
    using \Etr (left panel) and \Mc (right). The circles
    represent the experimental data for central 
    \XeSn collisions at 50~A\,MeV and the lines are the MMMC-NS  
    predictions with the prolate source. Full symbols and solid lines 
    represent the results for forward angles ($\theta_{\rm cm}\leq60^{\circ}$), 
    open symbols and dashed lines for sideward angles
    ($60^{\circ}<\theta_{\rm cm}\leq120^{\circ}$). Experimental 
    error bars are smaller than the symbol size.
    }
  \label{fig:Zdist_FS} 
\end{center}
\end{figure}

The angular anisotropy is also apparent in the yield distributions
as a function of the fragment atomic number $Z$ (Fig.~\ref{fig:Zdist_FS}).
More and larger fragments are produced in the forward region (polar angle
$\theta_{\rm cm}\leq60^{\circ}$) than at sideward directions ($60^{\circ}<\theta_{\rm cm}\leq120^{\circ}$),  
with a relative difference that increases with $Z$.
For central events selected on the basis of $M_{\rm c}$, 
the effect remains but the
multiplicities are slightly lower for large $Z$. On the other hand,
the multiplicities for \Zlcp in the forward direction are larger, 
presumably due to light particles emitted during the
pre-equilibrium phases of the reaction. In contrast, with the 
\Etr selection, the sideward multiplicities of light particles
increase noticeably and become equal to the forward  multiplicities.
This is obviously an auto-correlation produced by the selection method. 
Comparing the experimental multiplicities for \Zlcp
to the MMMC-NS calculation which are supposed to account for the 
equilibrium phase only,  
we conclude that nearly half of the measured light charged particles 
originate from either the early pre-equilibrium stage or from the 
expansion of the composite system prior to breakup 
(Fig.~\ref{fig:Zdist_FS} and Table~\ref{tab:table1}).

\begin{figure}
\begin{center}
  \epsfysize=6.0 cm
  \epsffile{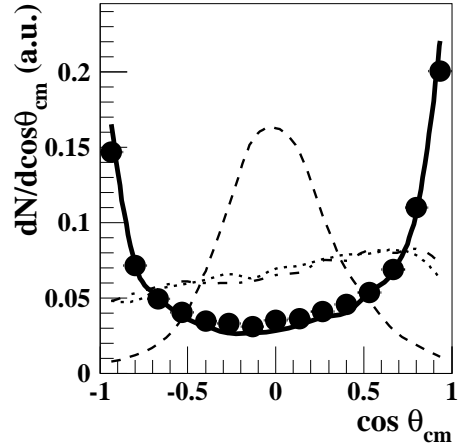}
  \caption{ 
    Angular distribution of the largest fragment ($Z_{\rm max1}$) 
    within an event. 
    Experimental data for central \XeSn collisions at 50~A\,MeV 
    (circles) and MMMC-NS predictions after filtering for the
    prolate (full line), spherical (dotted) and oblate (dashed)
    sources, and for the spherical source with non-isotropic flow 
    (dashed-dotted).
    All distributions are normalized to the same area. 
    The forward-backward
    asymmetry is caused by the experimental acceptance.
  }
  \label{fig:angdist_Zmax}
\end{center}
\end{figure}

As demonstrated in Fig.~\ref{fig:Zdist_FS}, 
the simulation with the prolate source reproduces the experimental 
forward-sideward difference in fragment yields. The spherical source 
model produces no angular dependence, as expected,
whereas the oblate source predicts an opposite trend with smaller yields in 
forward and backward directions. 
Another observable with a  strong shape discrimination power is the 
angular distribution of the largest fragment 
(Fig.~\ref{fig:angdist_Zmax}). Again, a strong forward-backward 
focusing is observed and only the MMMC-NS prolate source can 
describe it. The spherical source 
with isotropic flow yields a flat distribution but with a weak slope 
caused by the simulated INDRA acceptance. 
The superposition of a non-isotropic collective flow 
does not significantly modify this result because of the radial nature 
of the flow and because of the small magnitude required in the present case.
A non-isotropic flow, therefore, cannot simulate the 
effects of a spatial source deformation on the observables studied so far.

\begin{figure}
\begin{center}
  \epsfysize=7.5 cm
  \epsffile{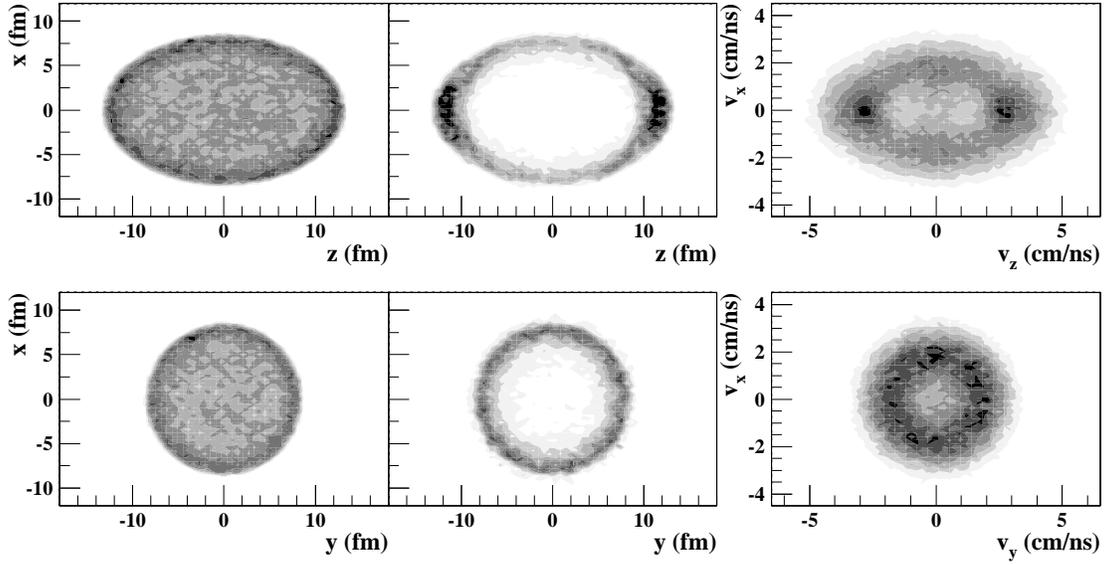}
  \caption{
    Two-dimensional distributions of fragments with $Z>4$, 
    calculated in the 
    MMMC-NS model with the prolate source and before filtering, 
    in coordinate space without Coulomb interaction (left), 
    with Coulomb interaction (center) and in 
    velocity space with Coulomb interaction (right). 
    The top panels show the 
    projections within a  centered longitudinal slice with widths 
    5~fm and 2~cm/ns in space and velocity, respectively. 
    The bottom panels show the corresponding projections,
    for transversal slices with the same thickness. 
    The $z$-axis is oriented in the beam direction, 
    the shading scale is linear.
  }
  \label{fig:rp_maps}
\end{center}
\end{figure}

In the MMMC-NS prolate simulation, the strong forward-backward 
focusing of the heaviest fragments originates from their particular 
location in the freeze-out volume. This is a consequence 
of the minimization of the Coulomb energy on an event-by-event
basis which favours a concentration of large fragments near the tips of 
the source and reduces the weight of partitions with high 
Coulomb energy, as, e.g., with heavy fragments 
compactly distributed near the center of the freeze-out volume.
As illustrated in  
Fig.~\ref{fig:rp_maps} (middle panels),
the transversal and longitudinal projections of the 
yields of heavy fragments ($Z>4$) for slices in the coordinate space 
show that the interior of the source is strongly depleted. 
Heavy fragments  
are emitted predominantly from the tips of the ellipsoid close to 
the surface. This focusing is
enhanced for larger fragment charges but is absent for light 
charged particles. Since  the radial collective flow is
directly correlated with the spatial fragment distribution 
(see Eq.~\ref{eq:vcoll}), the geometry of the source 
and its ring structure in the transversal cut
are reflected in the velocity space, as shown in 
the right panels of Fig.~\ref{fig:rp_maps}.
Depending on the relative strengths of the thermal and collective 
components, this space-momentum correlation is partly smeared out 
by thermal motion. 

An illustration of the decisive role of the Coulomb energy is given
in the  left panels of Fig.~\ref{fig:rp_maps}. If the Coulomb 
interaction is artificially switched off, heavy fragments are 
found to be much more homogeneously spread over the freeze-out volume.
There is still a slight concentration near the surface, caused by the 
geometrical advantage it gives for placing the other fragments 
of the partition.
In this case, the resulting angular distribution 
of heavy fragments becomes much flatter and does no longer agree 
with the experimental observation.

\subsection{Kinetic energies and the anisotropy 
  of the collective flow} \label{sec:f_dim}

The collective motion is not included in the energy balance of 
the MMMC-NS model. Therefore, the yield and charge distributions 
primarily discussed up to now are only weakly dependent on the strength 
of the included flow (via the applied filter).
Because of the radial form of the velocity profile (Eq.~\ref{eq:vcoll}),
also the fragment emission into a given angular interval and the fragment
angular distribution are 
hardly affected.
This is no longer valid if the fragment kinetic energies 
are considered since they depend directly on the strength of the 
considered flow. The observed anisotropies of the kinetic energy distributions
may be accounted for with either an angle-dependent strength parameter 
$V_{\rm coll} (\theta_{\rm cm})$ or, for a non-spherical source, also by adjusting 
the exponent $\alpha_{\rm coll}$.

\begin{figure}
\begin{center}
  \epsfysize=6.0 cm   
  \epsffile{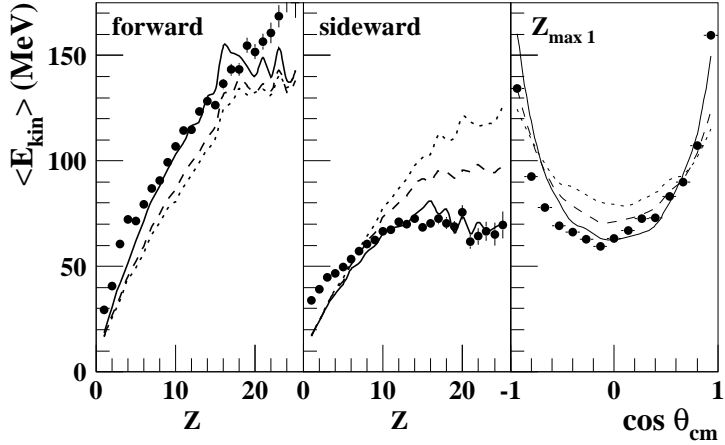}
  \caption{
    Left: Mean kinetic energy of fragments in the center-of-mass 
    as a function of $Z$ at forward angles ($\theta_{\rm cm}\leq60^{\circ}$), 
    calculated within the MMMC-NS model and after filtering,
    for the  prolate source with scaling exponents
    \ac = 0.5 (dotted), 1 (dashed) and 2 (full 
    lines). The symbols represent the experimental data for 
    central \XeSn collisions at 50~A\,MeV.
    Middle: The same as in the left panel but for sideward angles 
    ($60^{\circ}<\theta_{\rm cm}\leq120^{\circ}$). 
    Right: The mean kinetic energy in the center of mass 
    of the largest fragment ($Z_{\rm max1}$)
    as a function of the emission angle. 
  }
  \label{fig:alpha_dep} 
\end{center}
\end{figure}

The latter statement is illustrated in Fig.~\ref{fig:alpha_dep} (left and 
middle panels). The calculated mean kinetic energy 
as a function of $Z$ for the forward and sideward angular
regions depends significantly on the exponent \ac of the velocity profile
while the angle-independent $V_{\rm coll}$, adjusted to reproduce the experimental 
$\Sigma E_{\rm kin} (Z>5)$, varies little with $\alpha_{\rm coll}$. The anisotropy of the 
kinetic  energy, caused by the source elongation, is
strongly enhanced with larger $\alpha_{\rm coll}$. 
The same property is also reflected by the angular dependence 
of the mean kinetic energy of 
the largest fragments (Fig.~\ref{fig:alpha_dep}, right panel). 
In the model description, these fragments are mostly placed far from the 
center of the source and, therefore, have the
highest sensitivity to the exponent of the flow profile. 
The MMMC-NS prolate simulation gives the best agreement with the experimental 
data for $\alpha_{\rm coll}=2.0\pm0.2$ (statistical uncertainty). 
The sensitivity of $E_{\rm kin}(Z)$ to \ac is very
high as shown in Fig.~\ref{fig:alpha_dep}.
The difference between forward and sideward kinetic energies for
fragments with, e.g., $Z=20$ is $\approx75$~MeV in the experiment and for \ac=~2 but 
only 50~MeV and 25~MeV for \ac=~1 and 0.5, respectively.

\begin{figure}
\begin{center}
  \epsfysize=6.0 cm   
  \epsffile{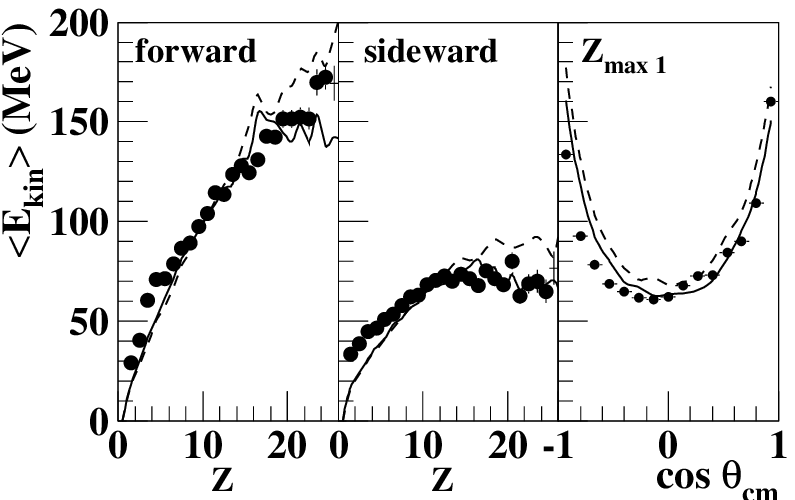}
  \caption{
    Mean kinetic energy of fragments in the center-of-mass for 
    central \XeSn collisions at 50~A\,MeV in the  
    representation of Fig.~\protect\ref{fig:alpha_dep}. 
    The symbols represent 
    the experimental data and the full lines are the MMMC-NS 
    model results for the prolate source with isotropic flow 
    and \ac = 2, as shown in Fig.~\protect\ref{fig:alpha_dep}. 
    The dashed lines represent the results for the prolate 
    source with non-isotropic flow of elongation 0.70:1.
  }
  \label{fig:prol_prol}
\end{center}
\end{figure}

Naturally, the forward-backward enhancement of the kinetic energies 
may also be very satisfactorily accounted for with a non-isotropic 
strength parameter. For \XeSn at
50 A\,MeV, the present analysis with the spherical MMMC source 
requires a large elongation of 0.3:1 (Table~\ref{tab:table2}, 
Spherical+e.f.).
With a slightly different procedure based on the SMM, 
this has also been demonstrated by Lavaud \textit{et al.}
for the \AuAu data from the present experiment 
\cite{lavaud_berkeley,lavaud}. 
The required effective elongation in velocity space 
ranges from $\approx$ 0.5:1 for \AuAu at 40 A\,MeV to $\approx$ 0.7:1 for the same 
reaction at 100 A\,MeV incident energy. 
For the prolate MMMC-NS source and \XeSn at 50 A\,MeV, 
the required elongation in velocity space 
is smaller, as expected. With the same source elongation
0.70:1 that was obtained for the coordinate space 
(Table~\ref{tab:table2}, Prolate+e.f.), an equivalent reproduction of 
the kinetic energies is obtained, even though the values for the 
largest $Z$ may be slightly overpredicted (Fig.~\ref{fig:prol_prol}).
This coincidence is interesting as it permits the source deformation 
to be explained rather naturally as resulting from a non-isotropic 
collective expansion.

\begin{figure}
\begin{center}
  \epsfysize=9.0 cm          
  \epsffile{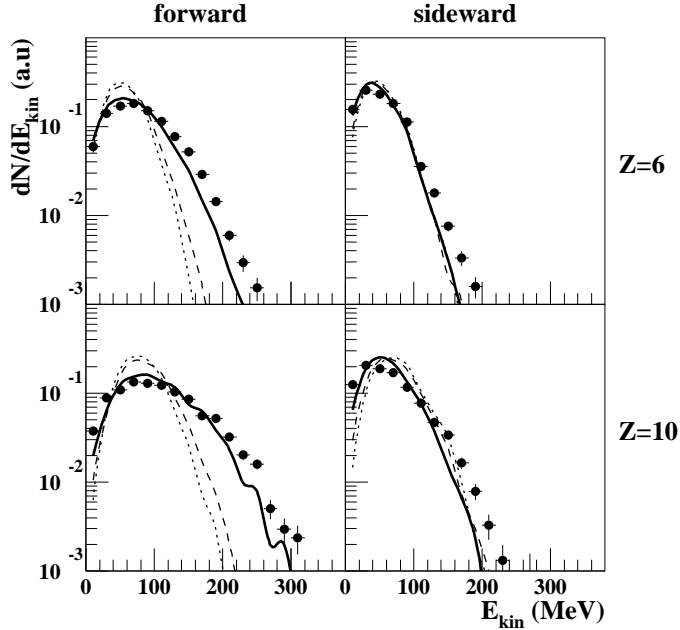}
  \caption{
    Kinetic energy spectra (in the center of mass) 
    of fragments with $Z=6$ (top) and $Z=10$
    (bottom)  at forward angles  (left panels) and at sideward
    angles (right panels). The circles are the experimental data for 
    central \XeSn collisions at 50~A\,MeV. The
    lines represent MMMC-NS predictions for the prolate source with
    scaling exponents \ac = 0.5 (dotted lines),
    1 (dashed lines) and 2 (full lines). 
  }
  \label{fig:Ek_Z6-10}
\end{center}
\end{figure}

The collective flow, on average, provides 
more than half of the fragment kinetic energy. It is then not 
surprising that the flow parameterization influences significantly 
also the shape of the energy spectra. This
is illustrated in  Fig.~\ref{fig:Ek_Z6-10} 
where the measured kinetic energy spectra of fragments 
with $Z=6$ and 10 are compared with the MMMC-NS prolate predictions 
for isotropic \vc and variable $\alpha_{\rm coll}$. There is little sensitivity
to \ac at sideward angles but, at the forward angles, the 
shapes of the experimental spectra are only reproduced with $\alpha_{\rm coll} \approx 2$.
At these angles of maximal spatial extension of the source, 
a velocity profile with \ac=~2 
results in broader spectra and higher kinetic energies than obtained
with the standard Hubblean profile.

\subsection{Global event-shape observables} \label{sec:e_shape_var}

In order to confirm the spatial and flow 
properties of the prolate source, predicted by the MMMC-NS 
model and determined in the preceding sections,
we shall now study the global event-shape observables sphericity $s$, 
coplanarity $c$ and flow angle $\Theta_{\rm flow}$ 
which refer to the event shape in momentum space.

Following Ref. ~\cite{cugnon}, we perform 
the event-by-event global analysis based on the three-dimensional 
momentum tensor, calculated in the center of mass of the reaction:
\begin{eqnarray}
\label{tens}
Q_{\displaystyle ij} = \sum^N_{\nu =1}
 \frac{ 1}{2m_{(\nu ) }} \ \  p^{(\nu )}_i \ p^{(\nu )}_j ~ \ ,
\end{eqnarray}
where $p^{(\nu )}_i$ is the $i$-th Cartesian coordinate ($i=1,2,3$) of the
center-of-mass momentum $p^{(\nu )}$ of the fragment $\nu$ with the 
mass $m_{(\nu )}$. With the chosen weighting $1/2m_{(\nu )}$, it represents the kinetic energy flow within the event.
To reduce the effects 
of secondary decay and of pre-equilibrium particle emission, the
 light charged particles are excluded and
the sum in Eq. \ref{tens} is taken only over
fragments with $Z\geq3$.  The  tensor $Q_{ij}$ can be represented 
as an ellipsoid to be described by three axes and by an orientation
defined by three angles in the three-dimensional momentum space. 
This is usually done by referring to the eigenvalues
$0\leq\lambda_1\leq\lambda_2\leq\lambda_3$ 
of the tensor $Q_{ij}$, with the normalization $\lambda_1+\lambda_2+\lambda_3=1$, 
and to the Euler angles defining the
eigenvectors ${\vec e}_1,{\vec  e}_2, {\vec e}_3$. 
From various possible combinations of
these parameters defining global variables \cite{cugnon},
we consider here the sphericity,
${\it s} = (3/2) (1-\lambda_3)$, which varies from 0 for a pencil-like event to 1 for a spherically symmetric event, 
the coplanarity, ${\it c} = ({\sqrt{3}}/2) (\lambda_2-\lambda_1)$, 
which extends from 0 (for spherical or pencil-like configurations) to 
$\sqrt{3}/4$ (for a disk),
and the flow angle $\Theta_{\rm flow}$ 
defined as the angle between ${\vec e}_3$
and the $z$-direction (the beam direction). The flow angle is most
useful for prolate shapes for which it describes the degree of alignment with respect to the beam axis ($\Theta_{\rm flow}=0^{\circ}$ for complete 
alignment).

\begin{figure}
\begin{center}
  \epsfysize=6.0 cm
  \epsffile{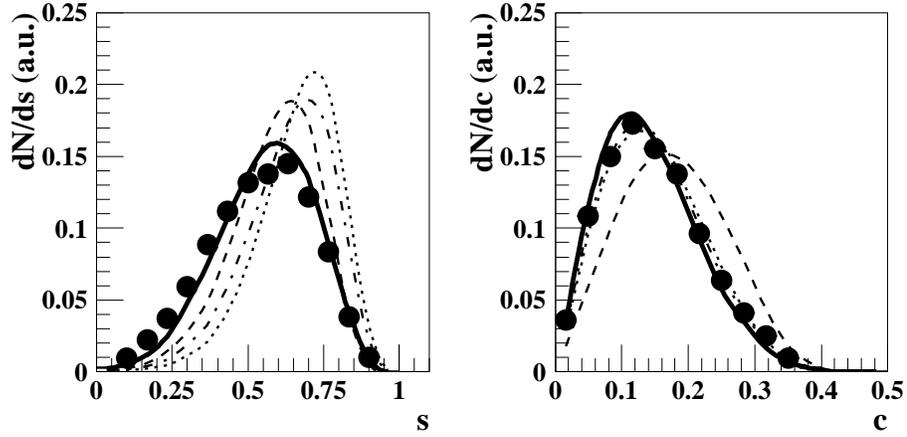}
  \caption{
    Sphericity (left panel) and coplanarity (right panel) distributions.
    The circles represent the experimental data for central \XeSn
    collisions at 50~A\,MeV and the lines are the
    MMMC-NS model predictions for the prolate (full line), 
    spherical (dotted) 
    and oblate (dashed) sources and for the spherical source with 
    non-isotropic flow (dashed-dotted).
    The distributions are normalized to the same area.
  }
  \label{fig:sph_copl} 
\end{center}
\end{figure}

\begin{figure}
\begin{center}
  \epsfysize=6.0 cm        
  \epsffile{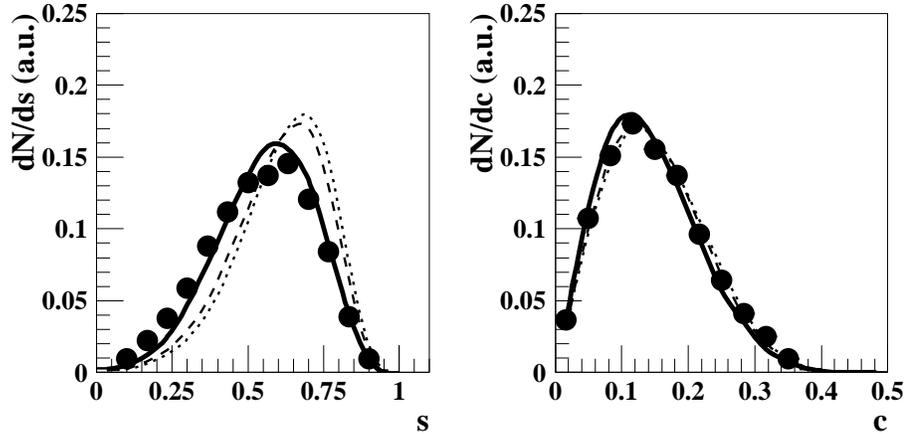}
  \caption{
    Sphericity (left panel) and coplanarity (right panel) distributions
    compared to MMMC-NS model predictions for the prolate source with 
    exponents of the flow profile  
    \ac = 0.5 (dotted), 1 (dashed) and 2 (full lines).
  }
  \label{fig:sph_copl_alpha} 
\end{center}
\end{figure}

Although these event-shape variables 
cannot be used to determine independently the source 
elongation and the flow profile, they can help us to confirm 
our finding with respect to the source parameters.
First, looking at the mean values of the sphericity and coplanarity 
given in Table~\ref{tab:table1}, we observe a good 
agreement between the experimental data and the MMMC-NS prolate
simulation.
Considering now the $s$ and $c$ distributions, depicted 
in Fig.~\ref{fig:sph_copl}, we note that the agreement with the 
experiment is remarkable. 
For comparison, other MMMC-NS shape configurations are included. 
By construction, the coplanarity distribution allows to exclude 
an oblate source only and does not discriminate between a spherical
form and a prolate one: both give close values of the coplanarity $c$. 
On the other hand, the sphericity distribution can exclude all 
shapes except the prolate one. This is valid 
even for the spherical shape with
non-isotropic flow, even though the latter properly describes 
the kinetic-energy anisotropy.  
A similar analysis for different values of the
exponent of the flow profile, presented for the prolate 
shape in Fig.~\ref{fig:sph_copl_alpha}, confirms that \ac=~2
best describes the distributions of fragment momenta 
in the present reaction. The parameter
\ac has quite a strong effect on the sphericity distribution, because
it enhances (for $\alpha_{\rm coll}>1$) or reduces 
(for $\alpha_{\rm coll}<1$) the elongation of a system in the momentum space,
 as compared to the standard Hubblean radial flow. 
In contrast, as expected, \ac influences only weakly the coplanarity which 
measures the azimuthal asymmetries of the prolate events.

\begin{figure}
\begin{center}
  \epsfysize=6.0 cm
  \epsffile{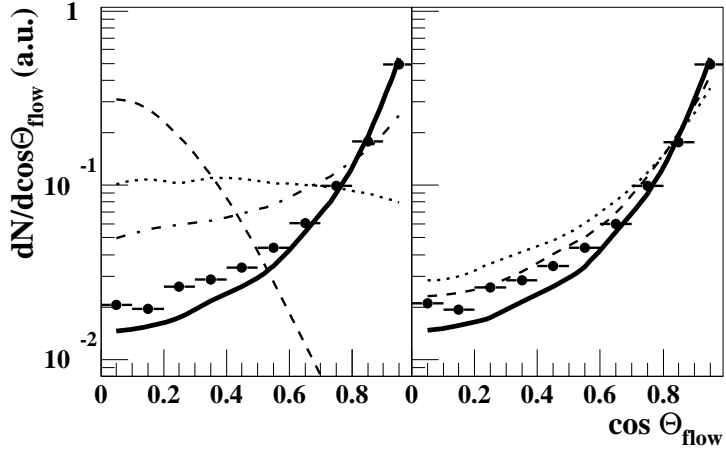}
  \caption{
    Flow angle distributions: 
    Experimental data for central \XeSn collisions at 50~A\,MeV
    (full circles) and MMMC-NS model predictions (lines)
    for different source shapes (left panel, notation as  
    in Fig.~\protect\ref{fig:Zmax_vs_theta}), 
    and for the prolate source
    with different exponents \ac = 0.5 (dotted), 
    1 (dashed) and 2 (full lines) 
    of the flow profile (right panel).
    The distributions are normalized to the same area. 
  }
  \label{fig:flow_angle} 
\end{center}
\end{figure}

A similar comparison has been done for the flow angle distribution 
(Fig.~\ref{fig:flow_angle}). It is strongly concentrated at small angles,
corresponding to the forward-backward enhancement of heavy fragments
(Fig.~\ref{fig:angdist_Zmax}) 
which contribute with large weights to the momentum tensor.
This behavior is correctly described only 
by the MMMC-NS prolate simulation. The yield at high $\Theta_{\rm flow}$ is slightly too low but can be expected to increase if fluctuations around the mean source deformation are included. The spherical configuration with anisotropic flow also 
reproduces the general trend but the agreement
with the experiment is only qualitative. Like in the experimental data, 
MMMC-NS prolate simulations produce, with a small probability, events 
having high values of $\Theta_{\rm flow}$. They 
correspond to rare partitions with the heaviest fragment being 
emitted sidewards or to compact arrangements of fragments with near-spherical
energy-flow distributions. 
The comparison of the experimental data and the
MMMC-NS prolate predictions for various values of \ac is shown in
Fig.~\ref{fig:flow_angle} (right panel).
The sensitivity to \ac is not very strong but the
general trend that the forward-backward focusing of 
the momentum tensor increases with \ac can, 
nevertheless, be clearly seen.

\section{Application to a heavier system} \label{sec:AuAu}

\subsection{\AuAu at 60~A\,MeV} \label{sec:AuAu60}

Given the possibility that the forward-backward enhancement observed for the 
fragment emission in \XeSn might be related to a lack of mutual stopping of 
the collision partners, an investigation of the mass dependence seems strongly 
suggested. 
A more efficient stopping may result from the larger nuclear radii and 
masses in even heavier systems. 
The present study has, therefore, been extended to the
\AuAu system at the similar beam energy of 60~A\,MeV that was 
measured as part of the same experimental campaign of INDRA at GSI
\cite{lavaud_berkeley,lavaud}.

The same methods were used for selecting central event samples (\bbmax $\leq0.1$)
according to
either \Etr or \Mc (cf. Sect.~\ref{sec:data_sel}), and their mean global 
characteristics were determined. They are listed in Table~\ref{tab:table3}. 
As in the \XeSn case, the reaction channels are highly fragmented with 
multiplicities that, apparently, scale with the mass ratio 1.6 of the 
two systems. Accordingly, the mean charge $\langle Z_{\rm max1} \rangle$
of the largest fragment of a partition is nearly the same as for \XeSna,
and the next largest fragments have also very similar mean charges. 
The mean sphericity $\langle s \rangle = 0.66$ for \AuAu is larger than the corresponding
value $\langle s \rangle = 0.53$ for \XeSna. As it turns out, however, this difference
is not connected to a more compact source configuration for the heavier
\AuAu system but rather to the larger fragment multiplicities observed in 
this reaction. The calculated value $\langle s \rangle = 0.67$ (after filtering, 
Table~\ref{tab:table3}) is obtained with a MMMC-NS source with 
elongation 0.70:1, identical to that derived for \XeSn at 50~A\,MeV.

The same analysis protocol as for \XeSn was followed to determine the 
parameters of the equivalent MMMC-NS source, its charge and mass,
the excitation energy (from the charge partitions), and 
the strength of the collective flow (from $\Sigma E_{\rm kin}(Z>5)$).
Subsequently, the spatial source deformation was derived from
the fragment-charge anisotropy, and the exponent $\alpha_{\rm coll}$ 
adjusted to the anisotropy observed for the fragment kinetic energies.
The agreement obtained for the mean characteristics of this reaction
in comparison with the experimental data is very satisfactory 
(Table~\ref{tab:table3}). Only the charged-particle multiplicity is again
underpredicted as preequilibrium emissions are not accounted for.
The deduced parameters are listed in Table~\ref{tab:table4}.
Obviously, to describe central collisions of
\AuAu at 60~A\,MeV, a prolate source is required that is larger, roughly  
in proportion to the initial masses, but otherwise very similar to the 
source deduced for \XeSna.

\begin{table}
  \caption{
    Mean characteristics of central \AuAu collisions
    at 60~A\,MeV and the MMMC-NS model predictions for the prolate source 
    with parameters determined from the comparison with the
    experimental data and listed in Table~\protect\ref{tab:table4}.  
    The presentation is the same as in Table~\protect\ref{tab:table1}.
    }
\label{tab:table3}
  \begin{center}
  \begin{tabular} {l c c c}
    \hline
   Observables & \multicolumn{2}{c}{Experiment} & MMMC-NS \\ 
    \cline{2-3}
    & \Etr selection & \Mc selection & \\
    \hline
    \Mc & 50 (53) & 56 (55) & 33 (34) \\
    $M_{\rm frag}$ & 9.3 (10.9) & 11.0 (11.9) & 10.2 (11.3) \\
    $Z_{\rm max1}$ & 13.7 (16.1) & 13.3 (14.9) & 14.4 (15.7) \\
    $Z_{\rm max2}$ & 9.4 (11.8) & 9.4 (10.7) & 9.4 (10.1) \\
    $Z_{\rm max3}$ & 6.9 (9.3) & 7.2 (8.4) & 7.5 (8.1) \\
    $s$ & 0.63 (0.66) & 0.65 (0.66) & 0.66 (0.67) \\
    $c$ & 0.13 (0.12) & 0.12 (0.12) & 0.12 (0.11) \\
    $\Sigma Z (Z>2)$ & 62 (79) & 68 (78) & 66 (75) \\
    $\Sigma E_{\rm kin} (Z>2)$  & 994 (1168) & 1138 (1273) & 875 (994) \\
    $\Sigma E_{\rm kin} (Z>5)$  & 586 (777) & 627 (764) & 577 (718) \\
    \hline
  \end{tabular}
\end{center}
\end{table}

\begin{figure}
\begin{center}
  \epsfysize=6.0 cm
  \epsffile{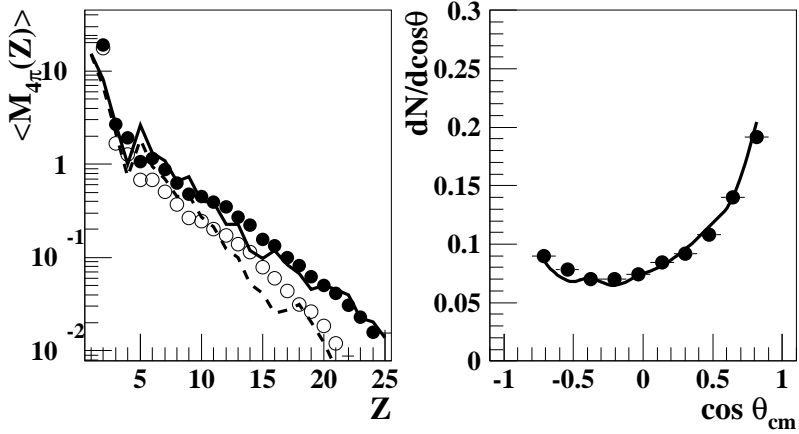}
  \caption{
    Left: Mean multiplicity of fragment charges, normalized to a
    solid angle of $4\pi$, with  selection of  central events
    using E$^{12}_{\perp}$. The circles represent
    the experimental data for central \AuAu collisions at 60~A\,MeV, 
    and the lines are the MMMC-NS model predictions calculated for the
    prolate source with parameters given in 
    Table~\protect\ref{tab:table3}. 
    The full circles and the solid line correspond to forward angles, 
    the open 
    circles and the dashed line correspond to sideward angles.   
    Right: Angular distribution of the largest fragment ($Z_{\rm max1}$) 
    within an event for the same reaction. 
    The full circles and the solid line represent
    the experimental data and the MMMC-NS model predictions, 
    respectively.    
    The distributions are normalized to the same area,
    the forward-backward
    asymmetry is caused by the experimental acceptance.
  }
  \label{fig:AuAu_Zdist_Zmax_theta} 
\end{center}
\end{figure}

\begin{figure}
\begin{center}
  \epsfysize=6.0 cm
  \epsffile{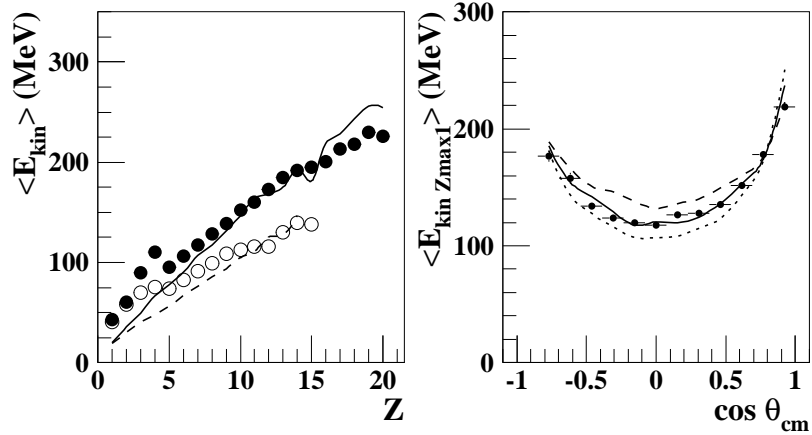}
  \caption{
    Left: Mean fragment kinetic energy as a function of $Z$ in the
    center-of-mass system. The  circles are experimental data for
    central \AuAu collisions at 60~A\,MeV and the lines are the 
    corresponding MMMC-NS prolate-source predictions for forward 
    (full circles and solid line) and sideward angles 
    (open circles and dashed line).
    Right: Mean kinetic 
    energy in the center of mass for the largest fragment ($Z_{\rm max1}$)
    as a function of the angle; the circles are the experimental data, 
    the dotted, full and dashed lines represent the MMMC-NS 
    prolate-source predictions with \ac~=~1, 1.5 and 2, respectively.
  }
  \label{fig:AuAu_Ek_vs_Z_theta} 
\end{center}
\end{figure}

The achieved description of the experimental $Z$ and $Z_{\rm max1}$ distributions 
and anisotropies is illustrated in Fig.~\ref{fig:AuAu_Zdist_Zmax_theta}.
The larger fragment yields at forward angles require the MMMC-NS source to 
be elongated in coordinate space. The observed anisotropy of the 
kinetic-energy distribution as a function of $Z$ and for the largest fragment 
as a function of angle are shown in Fig.~\ref{fig:AuAu_Ek_vs_Z_theta}. 
Also here a rather satisfactory agreement is obtained. In contrast to
\XeSna, a smaller flow exponent (\ac=~1.5) seems to be sufficient to account 
for the anisotropies in the isotropic non-Hubblean flow parameterization
(Fig.~\ref{fig:AuAu_Ek_vs_Z_theta}, right panel).

\subsection{\AuAu at 80 and 100~A\,MeV} \label{sec:AuAu_rest}

To extend the present analysis to energies 
beyond the Fermi energy domain is again of 
interest because the increasing phase space for Pauli-allowed nucleon-nucleon
collisions might enhance the mutual stopping of the colliding ions.
The data measured in the same experiment
for the \AuAu system at the higher energies 
80 and 100~A\,MeV are, in fact, supportive of such an interpretation.
The anisotropies of the fragment yields and energy 
distributions are slightly lower
at the highest bombarding energy. The dominating effects, however, 
are the rapid increase of the collective component of the fragment kinetic 
energy and the decreasing slopes in the fragment $Z$ distributions, known 
from the work of Lavaud \textit{et al.} \cite{lavaud_berkeley,lavaud}
and shown to continue to much higher bombarding energies by the FOPI
Collaboration \cite{reisdorf}. 

\begin{figure}
\begin{center}
  \epsfysize=5.0 cm
  \epsffile{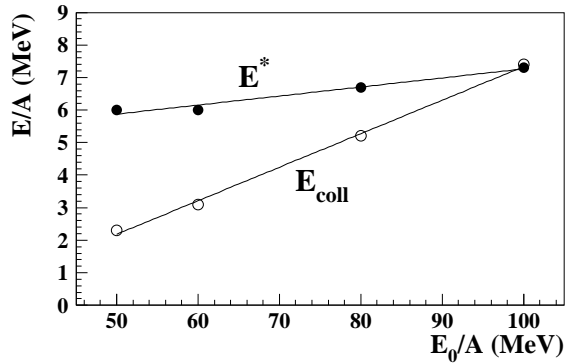}
  \caption{Mean thermal excitation energy
    (full circles) and collective flow energy (open circles)  at
    freeze-out, extracted by means of the MMMC-NS model, 
    as a function of the incident energy $E_0/A$ for 
    central collisions of \XeSn at 50~A\,MeV and \AuAu at 60,
    80 and  100~A\,MeV. The lines are linear fits, meant to 
    guide the eye.
  }
  \label{fig:synthesis}
\end{center}
\end{figure}

The obtained source parameters are listed in Table~\ref{tab:table4} and 
compared to those obtained at 60~A\,MeV for \AuAu and at 50~A\,MeV for
the \XeSn system. The charge of the source decreases slowly with 
rising beam energy, 
reflecting the decreasing part of the system that is emitted in the form 
of bound fragments.
The required thermal excitation energy rises slowly (by $\approx20\%$) 
while the collective energy increases rapidly, about
twice as fast as the available center-of-mass energy. 
The two quantities become approximately equal at a beam energy of 100~A\,MeV
(Fig.~\ref{fig:synthesis}).
The elongation of the source, on the other hand, remains 
remarkably stable with the system mass in the energy range
between 50~A\,MeV and 80~A\,MeV. At 100~A\,MeV, the source seems
to become slightly more compact. 

\begin{figure}
\begin{center}
  \epsfysize=5.0 cm
  \epsffile{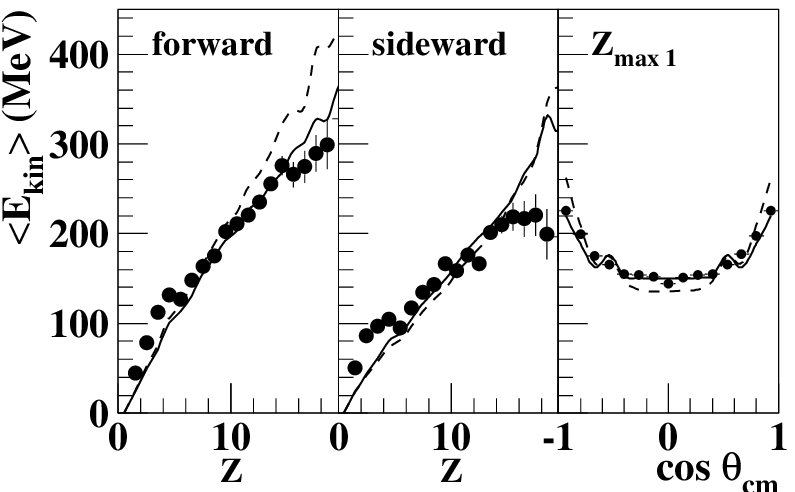}
  \caption{
    Mean kinetic energy of fragments in the center-of-mass for 
    central \AuAu collisions at 100~A\,MeV in the  
    representation of Fig.~\protect\ref{fig:prol_prol}. 
    The symbols represent 
    the experimental data and the full lines are the MMMC-NS 
    model results
    for the prolate source with isotropic flow and \ac = 1.2, 
    the dashed lines represent the results for the prolate 
    source with non-isotropic flow of elongation 0.76:1.
  }
  \label{fig:prol_prol_100}
\end{center}
\end{figure}

The anisotropy of the fragment kinetic energies decreases simultaneously with
a comparably slow pace. For the prolate source with isotropic flow, this 
appears in the form of a gradual decrease of the required flow exponent 
to $\alpha_{\rm coll}=1.2\pm0.1$ at 100~A\,MeV (Table~\ref{tab:table4}). 
For the spherical SMM source with ellipsoidal flow, the elongation in 
velocity space decreases from 0.5:1 to 0.7:1 \cite{lavaud_berkeley,lavaud}. 
A very satisfactory agreement is also obtained with the 
parameterization of the prolate MMMC-NS source and ellipsoidal flow if
the smaller elongation ${\mathcal R}= 0.76:1$ in coordinate space, required
at 100~A\,MeV, is also chosen for the collective velocities.
This is illustrated in Fig.~\ref{fig:prol_prol_100}, the kinetic energies 
for large $Z$ are again slightly overestimated (cf. Fig.~\ref{fig:prol_prol}).

\begin{table} 
  \caption{
     Incident energy $E/A$, collision energy in the center-of-mass system
     $E_{\rm cm}/A$ for the four studied reactions and characteristics of the 
     prolate freeze-out source in central collisions as obtained with the 
     MMMC-NS model. The \XeSn result of Table~\protect\ref{tab:table2} is
     included for easier comparison. 
    }
    \label{tab:table4}
  \begin{center}
  \begin{tabular} {l c c c c}
    \hline  
    Reaction & \XeSn & \multicolumn{3}{c}{\AuAu} \\
    \cline{3-5}
    $E/A$ (MeV)& 50 & 60 & 80 & 100 \\
    \hline
    $E_{\rm cm}/A$ (MeV) & 13.4 & 14.9 & 19.8 & 24.7 \\ 
    $Z_{\rm S}$ & 79 (76\%) & 125 (79\%) & 110 (70\%) & 95 (60\%) \\
    $A_{\rm S}$ & 188 & 312 & 275 & 238 \\
    $E^*$ (A\,MeV) & 6.0 & 6.0 & 6.7 & 7.3 \\
    $\mathcal{R}$ & 0.70:1 & 0.70:1 & 0.70:1 & 0.76:1 \\
     \vc (c) & 0.06 & 0.07 & 0.095 & 0.12 \\
     \ac & 2.0 & 1.5 & 1.3 & 1.2 \\
    $<E_{\rm coll}>$ (A\,MeV) & 2.3 & 3.1 & 5.2 & 7.4 \\
    \hline
  \end{tabular}
\end{center}
\end{table} 

\section{Summary and discussion.}  \label{sec:conclusions}

Angular anisotropies in the distributions of fragment size, 
fragment yield and fragment kinetic energy from  multifragmentation in
central \XeSn and \AuAu collisions have been measured with INDRA at GSI. 
The deformation extent of the phase-space available for the composite system 
at freeze-out, i.e. the spatial source elongation and the parameters 
of the collective flow, have been extracted from the experimental data
by means of the MMMC-NS model.
A good description is obtained with a prolate source shape in coordinate space,
oriented along the beam axis, and with an additional collective radial 
flow, superimposed on the thermal and Coulomb velocity components. 
With the usual assumption of an expanding spherical source 
the observed anisotropies cannot be explained. 
The additional assumption of a non-isotropic flow distribution can account for
the anisotropies of the kinetic energies but not for those of the fragment 
charge distributions.
The anisotropy of the fragment kinetic energies requires  
non-Hubblean flow profiles if an isotropic parameterization of the 
collective velocity component is chosen, in particular for the lower part 
of the studied range of bombarding energies. 
The magnitude of the flow increases rapidly with the beam energies 
and, for the heavier \AuAu system at 100~A\,MeV, assumes a nearly 
Hubblean profile.
 
An important role in the present statistical approach is
played by the Coulomb interaction which produces the anisotropies of
the fragment yield distributions. 
Under the constraint of total energy conservation, the  
position-dependent mutual Coulomb energies induce spatial correlations 
between the fragments. The superimposed radial flow
transforms  these spatial fragment-frag\-ment correlations into correlations
in their relative velocities. The heaviest fragments of the partition
were found to be a most sensitive probe of this scenario. This source 
of correlations contained in the individual 
fragment locations is lost if only the average Coulomb energy for a given partition is considered, as in the Wigner-Seitz approximation
used in standard versions of statistical multifragmentation models  
\cite{bondorf}. Coulomb induced spatial anisotropies 
are considered, besides in the MMMC, also in recent versions of the SMM model 
and found to be important in peripheral reactions of symmetric 
systems \cite{botvina}. 

Studying the influence of the system mass and  beam energy, 
we have observed that both \XeSn and \AuAu systems behave rather similarly. 
With increasing beam energy, the equilibrated fragment source
represents a decreasing fraction of the mass and charge of the collision system but maintains approximately the same prolate elongation oriented along the beam direction. At the highest energy of 100~A\,MeV, 
in the \AuAu system, the source elongation seems to be reduced. 
Interpreted as a result of the dynamics of the initial reaction stage,
it may indicate a more efficient mutual stopping 
of the colliding heavy ions with increasing beam energy.

At the lower bombarding energies 50 and 60 A\,MeV, the collective energy of
$\approx2$ A\,MeV constitutes only a small fraction of the total energy stored in 
the decaying system. It is, nevertheless, indispensable if the fragment 
kinetic energies are to be accounted for. Its magnitude has been deduced
rather consistently for various reactions in this energy range and with 
different methods \cite{bowman92,marie,bougault,steck96}.
The increase of the equilibrated excitation energy $<E^*>$ of the source with 
increasing beam energy is rather  small. In contrast, the collective flow 
energy $<E_{coll}>$ grows rapidly and becomes equal to the equilibrated
energy at 100~A\,MeV. 
It seems surprising that, even then, 
the complete decoupling of the statistical breakup from
the collective fragment motion should still be a viable assumption. 
It has been demonstrated that collective flow affects the partitioning of the 
system, in particular the survival probability for heavier fragments drops 
rapidly with increasing flow \cite{kunde95,pal,chikazumi,das,gulminelli}.
On the other hand, as shown in several of these references, the effect of 
the flow on the charge distributions may be simulated by increasing the 
value of the thermalized energy in the model description 
\cite{pal,gulminelli}. 
It is therefore quite likely that the flow effect is implicitly included in 
the parameters of the statistical description. At moderate flow values, 
the changes are expected to be very small \cite{das,gulminelli}.

A completely orthogonal approach to the question of the coexistence of 
equilibrated partitions and collective motion has recently been presented by 
Campi \textit{et al.}~\cite{campi}. 
Using classical molecular dynamics calculations and specific clustering 
algorithms, these authors find fragments to be preformed at the beginning of 
the expansion stage when the temperature and density are still high. 
The fragment charge distributions, reflecting the equilibrium at this early 
stage when the flow is still small, remain nearly unmodified down to the 
freeze-out density at which the flow has fully developed. While the initial 
compression density determines the collective flow observed asymptotically, 
the partitions are governed only by the initial system
 mass and its internal excitation energy. 
A fragment preformation at an early stage of the collision has been previously 
suggested also by other authors 
\cite{nebauer99,dani92,dorso95,barz96,puri96}.

Isotropic self-similar flow parameterizations were not found sufficient 
for the adequate description of the kinetic energies. Either a
non-Hubblean radial expansion with $\alpha_{\rm coll}>1$ or a non-isotropic strength 
distribution had to be considered, in particular in the range of lower 
bombarding energies. 
In ultra-relativistic heavy ion collisions,
expansion velocity profiles with $\alpha_{\rm coll} \simeq 2$ 
have been proposed to explain the transverse flow of hadrons
in a blast scenario \cite{lee2}. 
For the present regime, a microscopic explanation of this effect 
within the nonlocal BUU transport model has been
proposed by Morawetz \textit{et al.} and also identified
as a surface effect \cite{morawetz}. Even a decrease of the collective 
exponent with bombarding energy is predicted, in qualitative agreement 
with the present measurement. 
On the other hand, the need for $\alpha_{\rm coll}>1$ in the present case arises 
mainly from the large effect it has on the velocities of the heaviest fragments
which, in the MMMC-NS model, are placed at large distances from the center 
of the fragmenting source. In a dynamical scenario, these fragments 
would seem more likely to arise from colder parts of the residual projectile 
and target matter with incomplete dissipation of their initial energies, 
rather than from the rapid expansion of a surface shell. This does not 
reduce the importance of $\alpha_{\rm coll} \neq 1$ for successful parameterizations
of the fragment distributions and anisotropies (cf. Ref.~\cite{raduta}).

The anisotropy of the fragment kinetic energies has been equivalently well 
reproduced with an ellipsoidal flow distribution whose deformation was chosen
equal to the prolate source deformation in coordinate space. In an expansion 
scenario with a (non-isotropic) self-similar velocity profile, the shape
of the system will eventually be determined by the collective velocity 
distribution. The observed correlation of the deformation of the flow 
distribution with that required for the non-spherical source 
may provide the physical explanation for the latter and thus serve as 
an additional justification for the present statistical approach. 
The significance of this work is, independently, derived from the
consistent description of a large class of observables for conventionally 
selected central collisions. It demonstrates 
that the observed anisotropies do not necessarily imply 
the absence of equilibrium in the system. 
The constraining deformation parameters should be recovered from 
dynamical reaction models.

\ack{The authors would like to thank the staff of the GSI for providing high-quality \nuc{197}{Au} 
and \nuc{129}{Xe} beams and for technical support. 
Stimulating discussions with D.H.E. Gross and J.P. Wieleczko are gratefully acknowledged.
This work has been supported by the European Community under Contract No. ERBFMG\-ECT\-950083
and the CNRS-JINR Dubna agreement No 00-49, and by the IN2P3-CNRS, 
the DAPNIA/SPhN-CEA and the Regional Council of Basse-Normandie for the
construction of the INDRA detector. M.B. and C.Sc.
acknowledge the financial support of the Deutsche Forschungsgemeinschaft under
the Contract Nos. Be1634/1 and Schw510/2-1, respectively; D.Go. and C.Sf.
acknowledge the receipt of  Alexander-von-Humboldt fellowships.
}

\newpage


\newpage

\clearpage

\end{document}